\documentclass[]{aa}  

\usepackage{graphicx}
\usepackage{txfonts}
\usepackage{lipsum}
\usepackage{natbib}
\bibpunct{(}{)}{;}{a}{}{,} % to follow the A&A style
\usepackage[labelfont=bf]{caption}
\usepackage{booktabs}
\usepackage{array}
\usepackage{subcaption}         % necessary for continued figures, example in section 3
\usepackage{placeins}           % useful with \FloatBarrier, to keep 
\newcolumntype{P}[1]{>{\centering\arraybackslash}p{#1}}

\makeatletter
\AtBeginDocument{%
  \@ifundefined{linenumbers}{}{%
    \let\linenumbers\relax % 防止 linenoaa 在 \maketitle 里再打开
    \nolinenumbers
  }%
}
\makeatother

\begin{document}

   \title{On the dynamics, thermodynamics and fine structure of virtual erupting filaments}

%%%%%%%%%%%%%%%%%%%%%%%%%%%%%%%%%%%%%%%%
% Please do not include ORCIDs next to author names.
% Only ORCIDs authenticated by individual authors in EDP Sciences editorial system will be taken into account.
% ORCIDs included here will be removed.
%%%%%%%%%%%%%%%%%%%%%%%%%%%%%%%%%%%%%%%%

   \author{D. Donné\inst{1}
      \and Y. Zhou\inst{2}\fnmsep\thanks{Email: yuhaozhou@nju.edu.cn}
   \and H. Cremades\inst{3}
   \and R. Keppens\inst{1}
        }

   \institute{Centre for mathematical Plasma Astrophysics, Department of Mathematics, KU Leuven, Leuven, Belgium
   \and School of Astronomy and Space Science and Key Laboratory of Modern Astronomy and Astrophysics, Nanjing University, Nanjing, People's Republic of China
    \and Universidad de Mendoza, Mendoza, Argentina}

% \abstract{}{}{}{}{}
% 5 {} token are mandatory
 
  \abstract
   {It is not fully understood why some solar filaments erupt, while others do not. Those who do typically undergo a slow rise, followed by an acceleration phase, though this transition requires further investigation. Erupting prominences were observed to heat up during the acceleration phase but the origin of this heating remains unclear.  Moreover, some coronal mass ejections possess additional fine structure in white light observations, on top of the general three-structure morphology.}
   {We aim to elaborate on the dynamics of erupting prominences, investigate why erupting filaments heat up in the acceleration phase and correlate our findings with observations.} 
   {We employ the open-source software tool \texttt{MPI-AMRVAC} to solve the 2.5D MHD equations on a coronal domain that goes up to 300 Mm, using adaptive mesh refinement to attain a high resolution. We use controlled combinations of footpoint shearing and converging motions on an initial magnetic arcade, to get erupting flux ropes where self-consistent prominence and coronal rain formation occurs due to thermal instability. We find both non-erupting and erupting cases, as related to the energization of the system. We compare our erupting prominences with observations using data from the AIA Filament Eruption Catalog.}
   {{We find that the slow rise and impulsive phase of erupting prominences are modulated by magnetic reconnection. The transition from slow rise to acceleration is due to the transition from a low inflow Alfvén Mach number to a higher one.} For the first time, we demonstrate that {thermal conduction and compressional heating} can lead to prominence evaporation. We obtain clearly nested, circular fine structure in Extreme UltraViolet images of the ejected flux ropes, already present in their early evolution in the low corona. Some of this structure results directly from upwardly moving plasmoids interacting with the flux rope.}
   {We conclude that {thermal conduction and compressional heating} are highly relevant heating mechanisms in (erupting) flux rope interiors and that {magnetic reconnection dictates the entire early evolution of the CME: from the slow-rise phase to the impulsive phase.}}


   \maketitle

%%%%%%%%%%%%%%%%%%%%%%%%%%%%%%%%%%%%%%%%%%%%%%%%%%%%%%%%%%%%%%
\section{Introduction}
%%%%%%%%%%%%%%%%%%%%%%%%%%%%%%%%%%%%%%%%%%%%%%%%%%%%%%%%%%%%%%

    Solar prominences and filaments refer to the same coronal condensations, where density and temperature contrast to the ambient environment by around two orders of magnitude. It is now widely accepted that they are suspended by the magnetic field and that thermal instability plays a central role in their development \citep{parker1953,field1965}. A review on prominence formation, covering the theoretical insights from (up to current state-of-the-art) modeling efforts is provided in \citet{keppens2025}. Many numerical models exist in order to clarify their formation and further dynamics, such models include the levitation--condensation model \citep{kaneko2017,jenkins2021,brughmans2022,donne2024}, the evaporation--condensation model \citep{yuhao2014,xia2016,zhou2023,yoshi2025} and the plasmoid-fed condensation model \citep{zhao2022} which all incorporate thermal instability. Recently, a new flux-emergence-fed injection model has also been proposed by \citet{li2025}.

   Erupting prominences fall into the wider category of coronal mass ejections (CMEs). Their explosive behaviour affects interplanetary space weather and, if they collide with Earth, can influence the Earth's magnetosphere and interfere with satellites \citep{Tsurutani2019} as well as pose a serious health risk to astronauts during space flight missions \citep{dorman2008}. Hence, the ability to predict the (early) onset of CMEs is one of the outstanding open questions in heliophysics.

    Similarly to the theoretical models of solar prominences, many models exist to explain the nature of CMEs (see \citet{chen2011} and \citet{green2018} for an extensive list). Such models elaborate on two key aspects: the initial trigger and the further acceleration of an eruption. What all these models have in common is that the magnetic structure of CMEs consists of flux ropes as was already established by early models \citep{gibson1998,amari2000} and observations \citep{dere1999,hebe2006,vourlidas2014}. Flux ropes can only continue to erupt if its newly attained energy state at higher altitudes is lower \citep{forbes2000}. It is already well-accepted that shearing of the magnetic field is an important way of increasing the energy in flux ropes \citep[\& references therein]{chen2011,webb2012,georgoulis2019,green2018}. Recently, \citet{sen2025} conducted two simulations with initial non-force-free magnetic arcades that are initialized with a high and low shearing angle. They found that the simulation where the magnetic arcades had a high shearing angle form a flux rope and erupt shortly after, whereas the other simulation with an initially lower shearing angle did not. They concluded that shearing of the magnetic field lines is indeed vital to form flux ropes that erupt. The response of the flux rope to varying energy injections still needs further systematic investigations, and we will present one in this paper.

    Erupting prominences have also been observed to heat up during the impulsive phase of the CME \citep{landi2010, webb2012,lee2017}. However, any consensus on what causes this prominence heating is still absent. \citet{zaitsev2018} did an analytical investigation where the electric current flowing through the prominence is ad hoc increased and they hypothesized that the prominence \textit{can} be heated due to dissipation of this increased electric current. They justified this increase in electric current due to the magnetic Rayleigh-Taylor instability found in especially quiescent prominences \citep{berger2010,terradas2015,xia2016,kaneko2018,donne2024}. Heating related to magnetic processes of the erupting flux rope at low altitudes is also supported by \citet{landi2010}. They observed an erupting filament where the spectral properties of the filament changed from absorption to emission, indicating the heating of the prominence material. From their results, they ruled out CME heating mechanisms such as wave heating, thermal conduction and shock heating \citep{fillipov2002} and concluded that heating mechanisms from magnetic origin are more relevant, e.g. flows from magnetic reconnection, energetic particles induced by magnetic reconnection and other magnetic-related processes. Heating by flare-accelerated electrons has already been observed in a CME \citep{glesener2013}, but not (yet) for an erupting filament. Prominence evaporation has also been observed by \cite{wang2016}, though the cause of which was only speculated to be due to thermal conduction, changes in heating flux along the magnetic field and other processes. Hence, models elaborating on these (magnetic) heating mechanisms in the context of flux rope interiors are still heavily needed to explain this diversity in prominence behavior during a CME, such as its potential evaporation due to heating, its partial drainage and ejection.

    Three-part CMEs, consisting of a bright leading edge, cavity and bright core, are a recurring morphology in white light coronagraph observations \citep{illung1985}. In addition to the three-part morphology, additional, finer structures such as circular patterns are occasionally present \citep{dere1999,vourlidas2013}. Such fine structures are predominantly observed edge-on at or close to the solar limb, indicating that the line-of-sight is (close to) parallel to the flux rope axis \citep{hebe2004}. Hence, circular patterns are primarily interpreted in terms of the flux rope's helical magnetic field \citep{dere1999}. Though whether these fine structures already pre-exist before or during the impulsive phase of the CME in the low corona has not yet been looked into.

    In this paper, we extend our previous work \citep{donne2024} but now focus on erupting filaments in 2.5D. By conducting six simulations where the footpoint driving motion has a varying shear during flux rope formation, we examine the response of the flux rope to these drivers that differ in terms of the energy they inject into the system. We obtain both erupting- and non-erupting filaments. For all the erupting cases, the slow-rise phase of the CME is retrieved. {For the very first time, we provide a clear demonstration that thermal conduction and compressional heating can evaporate a prominence.} We synthesised our data and searched for observed CME morphologies that closely resemble our results. Hence, this paper is organised as follows. Section 2 provides methodological changes we made with respect to our previous work \citep{donne2024}. Section 3 exhibits our results with discussion. We end with a summary and conclusion of our findings.

\section{Methodology}
We use MPI-AMRVAC \citep{amrvac3.0} to solve the 2.5D resistive-MHD equations. Our numerical configuration is mostly similar to our previous work \citep{donne2024}, i.e. we reuse the same initial density and pressure distribution at an isothermal temperature $T_0=1 \ \mathrm{MK}$, {we use magnetic field splitting to solve the MHD equations where the magnetic field $\mathbf{B}$ is split into a time-independent $\mathbf{B}_0$ and time-dependent component $\mathbf{B}_1$ (see \citet{xia2018} for more details)} and we include the same source and sink terms such as gravity, radiative cooling, static background heating and anisotropic thermal conduction. Main differences have been made with dimension reduction from 3D to 2.5D, the simulation domain, initial magnetic field, resistivity and boundary conditions which will be elaborated upon now.

The first difference is the domain and magnetic configuration: a single arcade system has now been changed to a triple arcade system since a single arcade system is theorized to be unsuitable for CME driving  \citep{antiochos1999}. Hence, by extending the domain to $\vert x \vert \leqslant 30 \ \mathrm{Mm}$ a triple arcade system is retrieved. Further elongating the height $\vert y \vert \leqslant 300 \ \mathrm{Mm}$ allows for the study of CME acceleration before any interaction with the top boundary. The initial magnetic field is equal to
\begin{align}
B_x & =-\left(\frac{2 L_\mathrm{a}}{\pi a}\right) B_0 \cos \left(\frac{\pi x}{2 L_\mathrm{a}}\right) \exp \left[-\frac{y}{a}\right], \\
B_y & =B_0 \sin \left(\frac{\pi x}{2 L_\mathrm{a}}\right) \exp \left[-\frac{y}{a}\right], \\
B_z & =-\sqrt{1-\left(\frac{2 L_\mathrm{a}}{\pi a}\right)^2} B_0 \cos \left(\frac{\pi x}{2 L_\mathrm{a}}\right) \exp \left[-\frac{y}{a}\right],
\end{align}
{with $B_0=8\ \mathrm{G}$ the magnetic field strength} and $L_a = 20 \ \mathrm{Mm}$. The magnetic scale height has now been changed to $a=2\mathcal{H}_0$, with $\mathcal{H}_0 \approx 50 \ \mathrm{Mm}$ the pressure scale height. This results in a constant plasma beta value smaller than one at $t=0$ over the entire domain, i.e.
\begin{equation}
    \beta = \dfrac{p}{B^2 / (2\mu_0)} = \dfrac{2\mu_0p_0}{B_0^2} \exp\bigg[ -y \bigg(\dfrac{1}{\mathcal{H}_0} - \dfrac{2}{a}\bigg) \bigg]
\end{equation} 
with $p_0 = 0.3 \ \mathrm{erg} \ \mathrm{cm}^{-3}$. For $a>2\mathcal{H}_0$, $\beta$ would decrease with height, whereas in our case with $a=2\mathcal{H}_0$, it remains constant.

The resistivity in the present work is set to a uniform value of $\eta = 1.17 \cdot 10^{12} \ \mathrm{cm}^2 \, \mathrm{s}^{-1}$ or $\eta=0.001$ in code units. 

Lastly, we elucidate on the boundary conditions. For all boundaries, the magnetic field is calculated according to a second order zero gradient extrapolation. In the bottom boundary the gas pressure is extrapolated in a zero gradient, second order fashion as well and using the ideal gas law the density is calculated such that $T=1 \ \mathrm{MK}$. This implies that we are dealing with a purely coronal evolution, and we do not consider the coupling across the transition region to the chromosphere. Defining the (positive) shearing constant $\sigma$, the velocity footpoint motion consists of a converging component in the $x$-direction and shearing component in the $z$-direction $\mathbf{v} = \nu (\,\mathbf{\hat{i}} - \sigma\,\mathbf{\hat{k}})$. In this work we conduct six simulations for different $\sigma \in \{0, 0.75, 1, 1.25, 1.5, 2\}$ that continue until $t=t_1=5 \, 600$ s. In other words, simulations with larger $\sigma$ incorporate more shearing into the simulation domain and hence result in a situation where the flux rope is in a higher energy state. The factor $\nu$ is prescribed as follows:
\begin{equation}
   \nu  = 
        \begin{cases}
            -f(0, t_1, t)\,v_0 \sin\bigg(\frac{\pi x}{L_a}\bigg) \quad &\mathrm{if} \ t \leq t_1 \ \mathrm{and} \ \vert x \vert < L_a ,  \\
             0 \quad &\mathrm{otherwise},
        \end{cases}
\end{equation}
Note that at $t \leq t_1$ we only allow the central magnetic arcade $\vert x \vert < L_a$ to converge towards the central polarity inversion line (PIL) to form the central flux rope. Here, $v_0 = 2 \ \mathrm{km \, s^{-1}}$, which is relatively close to the observed photospheric flow values of $0.75 \pm 0.05 \ \mathrm{km \, s^{-1}}$ \citep{delmoro2007}. For all $\sigma$-cases, footpoint driving is turned off at $t_1=5\,600 \ \mathrm{s}$ and this happens smoothly according to the linear ramp function $f(0, t_1, t)$. It allows for a smooth transition of $t_\mathrm{ramp} = 1\,000 \ \mathrm{s}$ into and out of the state at $t=0$ and $t=t_1$. For general values, $f(t_\alpha, t_\beta, t)$ is defined as:
\begin{equation}
   f(t_\alpha, t_\beta, t)  = 
        \begin{cases}
            \dfrac{t - t_\alpha}{t_\mathrm{ramp}} \quad &\mathrm{if} \ t_\alpha \leq t < t_\alpha + t_\mathrm{ramp},  \\
            1 \quad &\mathrm{if} \  t_\alpha + t_\mathrm{ramp} \leq t < t_\beta - t_\mathrm{ramp},  \\
            \dfrac{t_\beta - t}{t_\mathrm{ramp}} \quad &\mathrm{if} \  t_\beta - t_\mathrm{ramp} \leq t < t_\beta, \\
            0 \quad &\mathrm{if} \ t=t_\beta.
        \end{cases}
\end{equation}
For the side boundaries we copy the density and pressure from the adjacent inner cell. The $x$-component of the velocity is set to zero $v_x=0$ whereas its other components are copied from the adjacent inner cell, depending on the time:
\begin{equation}
   v_{y, z}  = 
        \begin{cases}
            f(0,t_1, t) \, v_{y, z, \, \mathrm{adjacent}} \quad &\mathrm{if} \ t \leq t_1 \,,  \\
            f(t_1,\infty, t) \, v_{y, z, \, \mathrm{adjacent}} \quad &\mathrm{otherwise.} 
        \end{cases}
\end{equation}
{The time dependency is only due to the ramp function $f(t_\alpha, t_\beta,t)$, whose only purpose is to vary smoothly from symmetric- to static boundary conditions at the side boundaries and vice versa. More specifically,} at $t=t_1$ we desire the ghostcell values of $v_{y, z}$ to be zero again to damp any growing {(numerical) instability} during flux rope formation. For $t>t_1$ we gradually restart copying $v_{y, z}$ from the side boundaries again. Additionally, at all times for both side boundaries a maximum value of $6 \ \mathrm{km} \, \mathrm{s}^{-1}$ is imposed on the velocity $v_z$.

In the top boundary, the initial isothermal requirement $T=1 \ \mathrm{MK}$ is relaxed such that both the inner pressure and density are extrapolated to the ghostcells using the same second order zero gradient extrapolation as with the magnetic field. We augment this condition with the restriction $v_y \geqslant  0$ to prevent any inflow from occurring.  The velocities $v_x$ and $v_z$ are zero and for $v_y$ we copy the adjacent cell values with a maximum velocity cap of $0.05$ code units or $6 \ \mathrm{km} \, \mathrm{s}^{-1}$ for $t\leq t_1$. For $t>t_1$, the maximum cap for $v_y$ is disabled.

The base resolution is $144 \times 720$ with four AMR levels (one base grid and three refinement levels), reaching an effective resolution of $1\,152 \times 5 \, 760$. The bottom $y < 2 \ \mathrm {Mm}$ and top $y > 299 \ \mathrm{Mm}$ are maximally refined. In addition, locations $\vert x \vert \geq 27 \ \mathrm{Mm}$ are maximally coarsened. Where the $z$-component of the split-off magnetic field is $B_{1, z} < -1 \ \mathrm{G}$ (or $B_{1, z} < -0.5 \ \mathrm{code} \, \mathrm{units}$), the magnitude of the total current density $J \equiv \vert \mathbf{J} \vert > 4.76 \ \mathrm{statamp} \, \mathrm{cm}^{-2}$ (or $J > 1  \ \mathrm{code} \, \mathrm{unit}$) and local temperatures $T < 0.1 \ \mathrm{MK}$, the region is maximally refined. This results in an effective resolution of 52 km. 

\section{Results \& discussion}

\begin{figure}[thp]
\begin{center}
\includegraphics[width=0.49\textwidth]{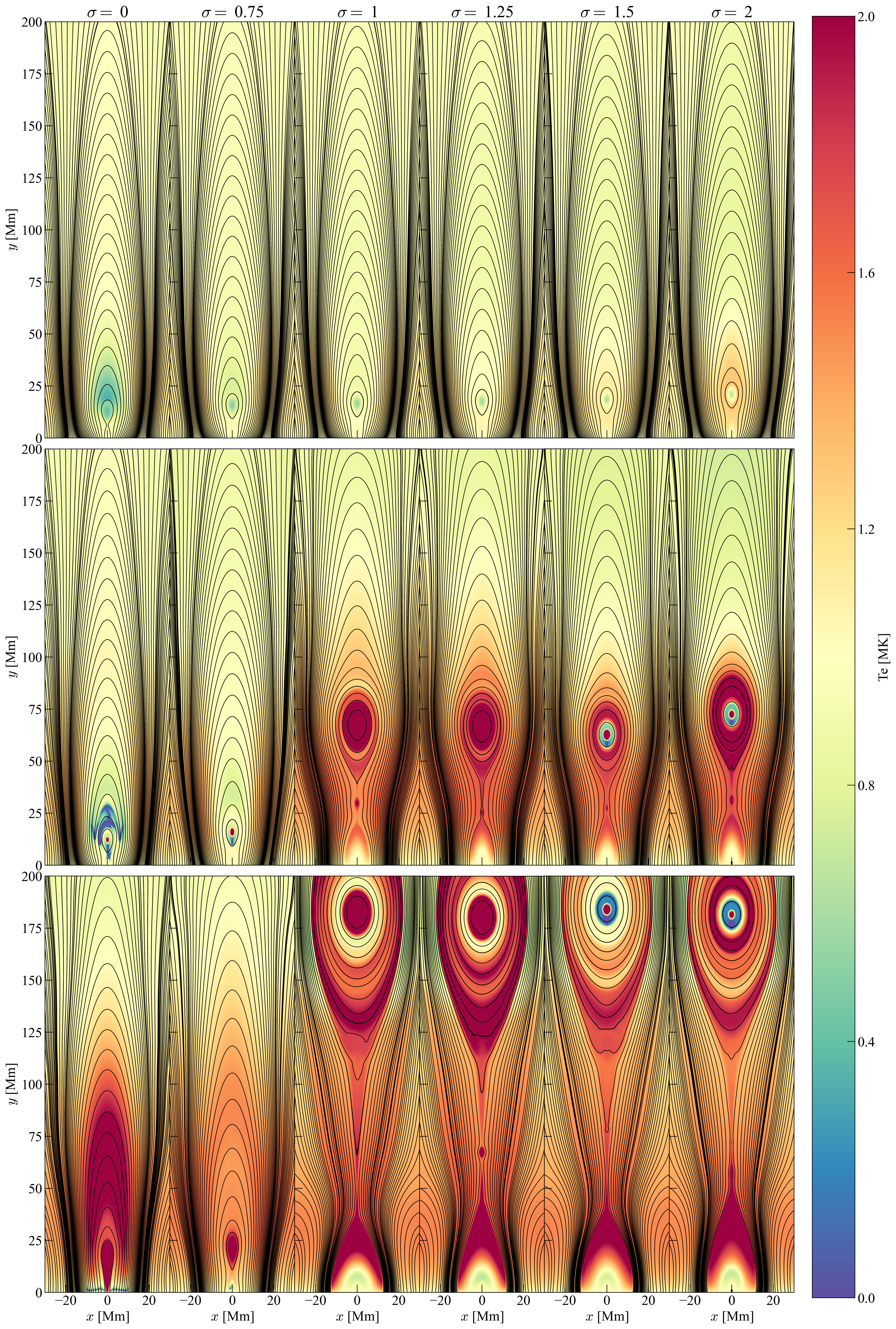}
\caption{Temperature evolution of the six simulations. From left to right, the columns are the simulations for $\sigma \in \{0, 0.75,  1, 1.25, 1.5,2 \}$. The first row are snapshots of the simulations at the time when footpoint driving motion is disabled $t=t_1$; the second row when coronal rain ($\sigma=0$) and the solar prominence ($\sigma=0$ and $0.75$) develop and plasmoids ($\sigma \in \{1, 1.25, 1.5,2\}$), and finally, the third row at the end of the respective simulations. The thin black lines show magnetic field lines. Note that the end stage is at different timestamps for the different $\sigma$ cases: $253 \ \mathrm{min} \  (\sigma \in \{ 0, 0.75, 1\}), \, 257 \ \mathrm{min} \ (\sigma=1.25), \, 234 \ \mathrm{min} \ (\sigma=1.5)$ and $ 232 \ \mathrm{min} \ (\sigma=2)$. An animation of this figure is available in the journal's online webpage.}\label{fig:te_evolution}
\end{center}
\end{figure}

We display three snapshots of the temperature of each simulation in Fig. \ref{fig:te_evolution}. The first row shows the system at the moment when footpoint driving motion has been turned off at $t=t_1$. In the second row, $\sigma=0$ and $\sigma=0.75$ are shown when the condensations form, but these two systems do not erupt. The snapshots of the other cases $\sigma > 0.75$ do erupt and are displayed at the time when their first plasmoid forms. In the third row the last snapshot of each simulation is displayed.

The qualitative results are as follows and can be seen in the provided animation of the figure. At $t=t_1$ (first row), it can already be seen that stronger shearing results in the flux rope being located higher. In addition, the surrounding environment of the flux rope heats up more if shearing is increased. In all cases, a solar prominence forms due to thermal instability and only in the case $\sigma=0$ does coronal rain form in addition.  This result is not surprising considering that the $\sigma=0$ case closely resembles our 3D simulated prominence. In the cases $\sigma=0$ and $\sigma=0.75$, the solar prominence escapes the flux rope due to mass slippage \citep{low2012,donne2024}, but the flux rope does not erupt. In all the other cases, a prominence eruption is retrieved but with added diversity: for $\sigma\in\{1,1.25\}$, the solar prominence evaporates during the onset of eruption and only for $\sigma \in \{1.5,2\}$ does the solar prominence remain stably in the flux rope for the entire CME evolution. 

This diversity in our results is visualised in the last row. Two cases ($\sigma=0$ and $\sigma = 0.75$) do not erupt, and all other four cases do. Two of the four erupting cases does not contain a prominence ($\sigma \in\{1,1.25\}$), while the others do ($\sigma \in\{1.5,2\}$), is in agreement with the finding of \citet{low2001}. They discussed that despite their optimal magnetic topology, flux ropes do not always contain a solar prominence due to the fact that flux ropes can have varying thermodynamic interiors. Furthermore, in all erupting cases we obtain plasmoids as well. Therefore, we will discuss the CME dynamics and energies, the varying thermodynamics of their flux rope interiors and lastly our synthetic CME morphologies and their agreement with observations. For all data shown, only the results are shown/used when the flux rope centre obeys $y_\mathrm{frc} \leqslant 200 \ \mathrm{Mm}$ unless specified otherwise.

\subsection{Dynamics of the CMEs}

\begin{figure}[thp]
    \includegraphics[width=0.48\textwidth]{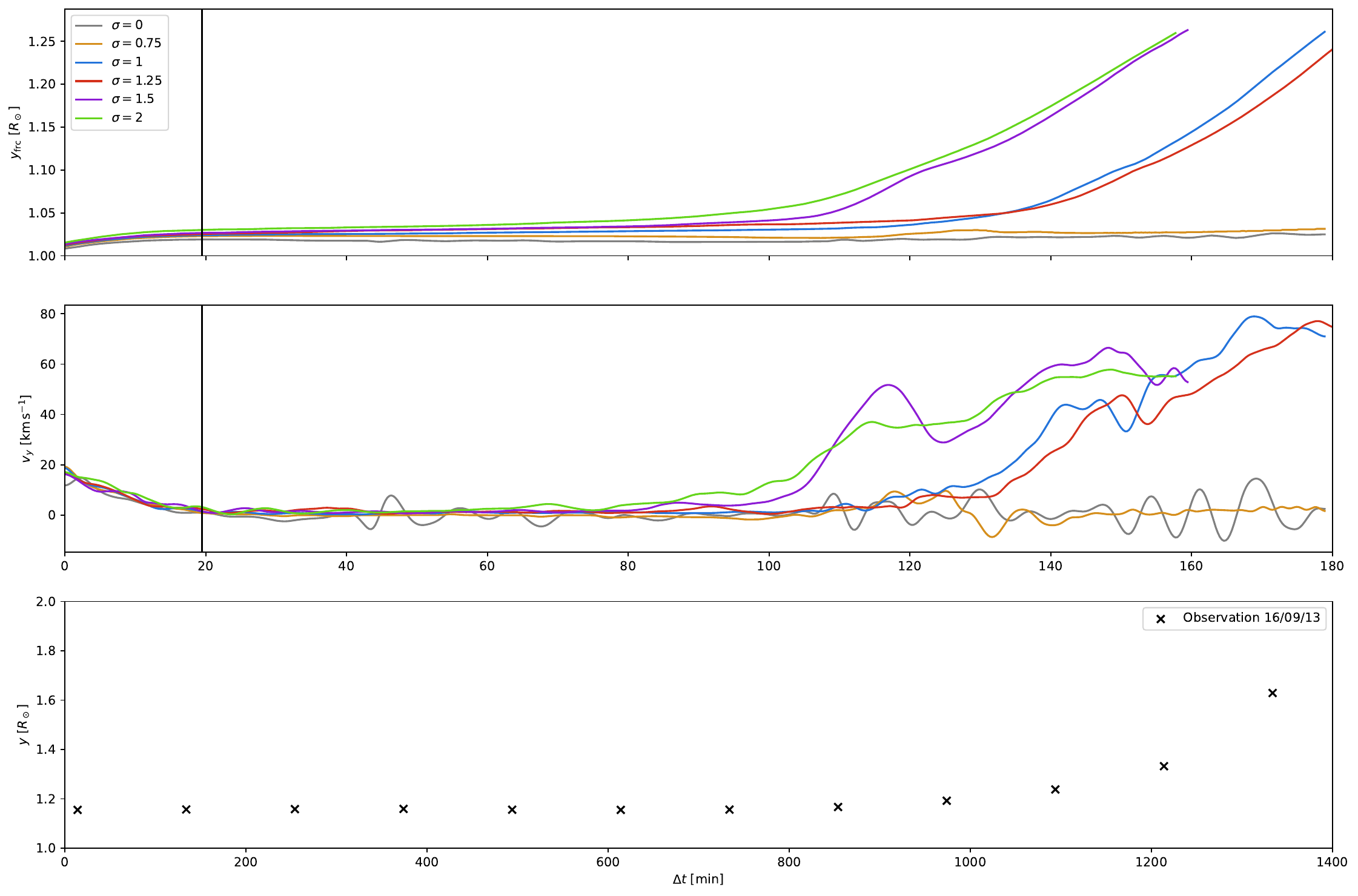}
    \caption{Height evolution $y$ of the flux rope centres in solar radius (top panel) and vertical velocity $v_y$ in km/s for the six different shearing cases. Note that $\Delta \, t=t - t_\mathrm{fr}$ indicates the time difference with respect to the moment when the flux rope has already been formed.  The black vertical line indicates the instant when footpoint driving motion has been disabled at $t=t_1$ or $\Delta \, t = t_1 - t_\mathrm{fr}$. The last point of all height- and velocity curves indicates the moment when the flux rope centre has crossed the altitude $y=200 \ \mathrm{Mm}$. Bottom panel displays the height evolution of an observed erupting prominence, data from \citet{DiLorenzo2025}.}\label{fig:height_evolution}
\end{figure}

By tracking the location of the flux rope centre, {which is defined as the (only) point with zero magnetic curvature within the flux rope,} we can obtain a height-time graph. This is provided by the first panel of Fig. \ref{fig:height_evolution}. The tracking of the flux rope centre starts at $t_\mathrm{fr} \approx 74 \  \mathrm{min}$ for all cases and so the height is plotted in function of $\Delta \, t = t - t_\mathrm{fr}$. The black line indicates the time at which footpoint motion is disabled. It can be seen that there is {almost} a strict ordering, i.e. higher sheared flux ropes are strictly located higher at each time $t$ {except for $\sigma=1.25$ which gets overtaken by $\sigma=1$}. The case $\sigma=0$ did not erupt nor does it give any sign of eruption and hence maintains the same altitude. The other case $\sigma=0.75$ did not erupt either within our simulation time, but the altitude of its flux rope centre has increased from $16 \ \mathrm{Mm}$ at the disabling of the footpoint motion to $20 \ \mathrm{Mm}$ at the end of its simulation. The other cases did erupt within the simulation time and display an increasing curve as expected. For the erupting cases $\sigma \in \{1, 1.25, 1.5,2\}$, we see that after the black line, the flux rope very slowly increases in height, also known as the slow-rise phase of the CME, and after some time it does so impulsively. This impulsive acceleration corresponds to the onset of higher magnetic reconnection rates \citep{priest1986, shibata2001, zhao2017}.

\begin{figure}[thp]
    \includegraphics[width=0.48\textwidth]{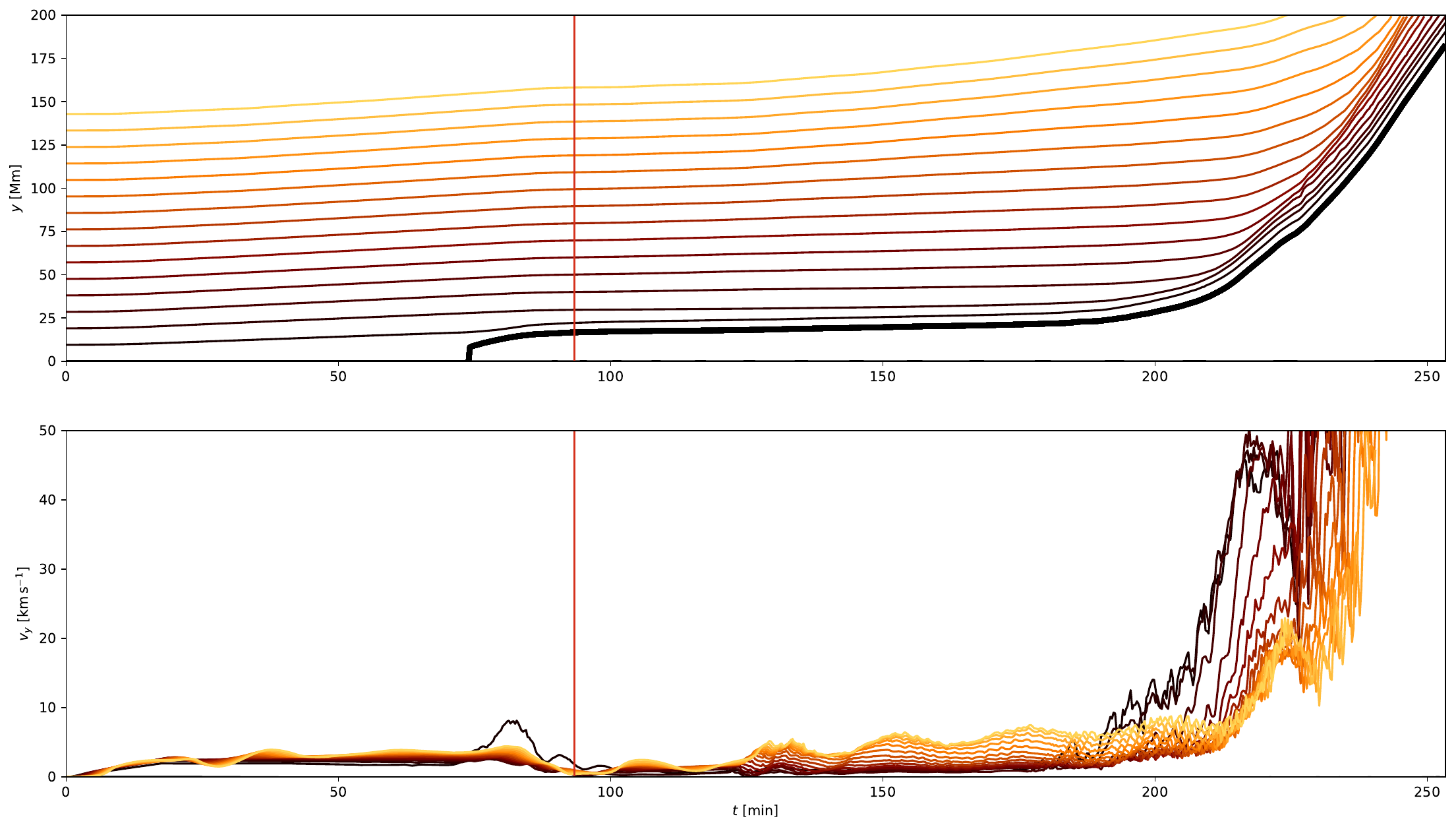}
    \caption{Height $y$ in megameters (top panel) and vertical velocity $v_y$ in kilometers per second (bottom panel) of the central arcades' apexes in function of time $t$ in minutes for $\sigma=1$. The change in colours ranging from black to orange indicate lower lying arcades to higher located arcades, respectively. The thick black line is the evolution of the flux rope centre. The vertical red line marks the time when footpoint driving is disabled at $t=t_1$.}\label{fig:arcade_expansion}
\end{figure}

Recently, the works of \citet{liu2025} and \citet{xing2024b} demonstrated that the slow-rise phase of the overlying strapping field is due to continuous shearing of the magnetic field from footpoint driving motions. Since we disable footpoint driving motions some time after the formation of the flux rope, we expect no expansion of the overlying arcades from their findings. To compare our results with theirs, we retrieve the location of the apexes of the magnetic arcades and their associated vertical velocities as follows. Each magnetic field line is computed by following its seedpoint which co-moves with the local velocity field. When the magnetic field lines are traced from their seedpoints, their apexes are easily obtained by locating the global maximum of the magnetic field lines. The corresponding vertical velocities of the apexes are then acquired from the local values of the velocity field at the location of the apexes. Fig. \ref{fig:arcade_expansion} shows the height evolution (top panel) of the apexes of our central arcades for the $\sigma=1$ simulation which is in addition complemented with the corresponding vertical velocity of the apexes (bottom panel). Our other erupting simulations $\sigma \in\{1.25, 1.5,2\}$ show exactly the same behaviour, i.e. there is no (significant) expansion of the strapping arcades. Indeed, vertical velocities for all arcades for all cases of $\sigma$ display values $v_y < 10 \, \mathrm{km} \, \mathrm{s}^{-1}$. 
%Therefore, our results agree with those of \citet{xing2024b} and \citet{liu2025}.
{Therefore, our results are consistent with the interpretation of \citet{xing2024b} and \citet{liu2025}, in the sense that without continuous footpoint driving, no significant expansion of the overlying arcades is expected.}

\begin{figure}[thp]
    \centering
    \includegraphics[width=0.98\linewidth]{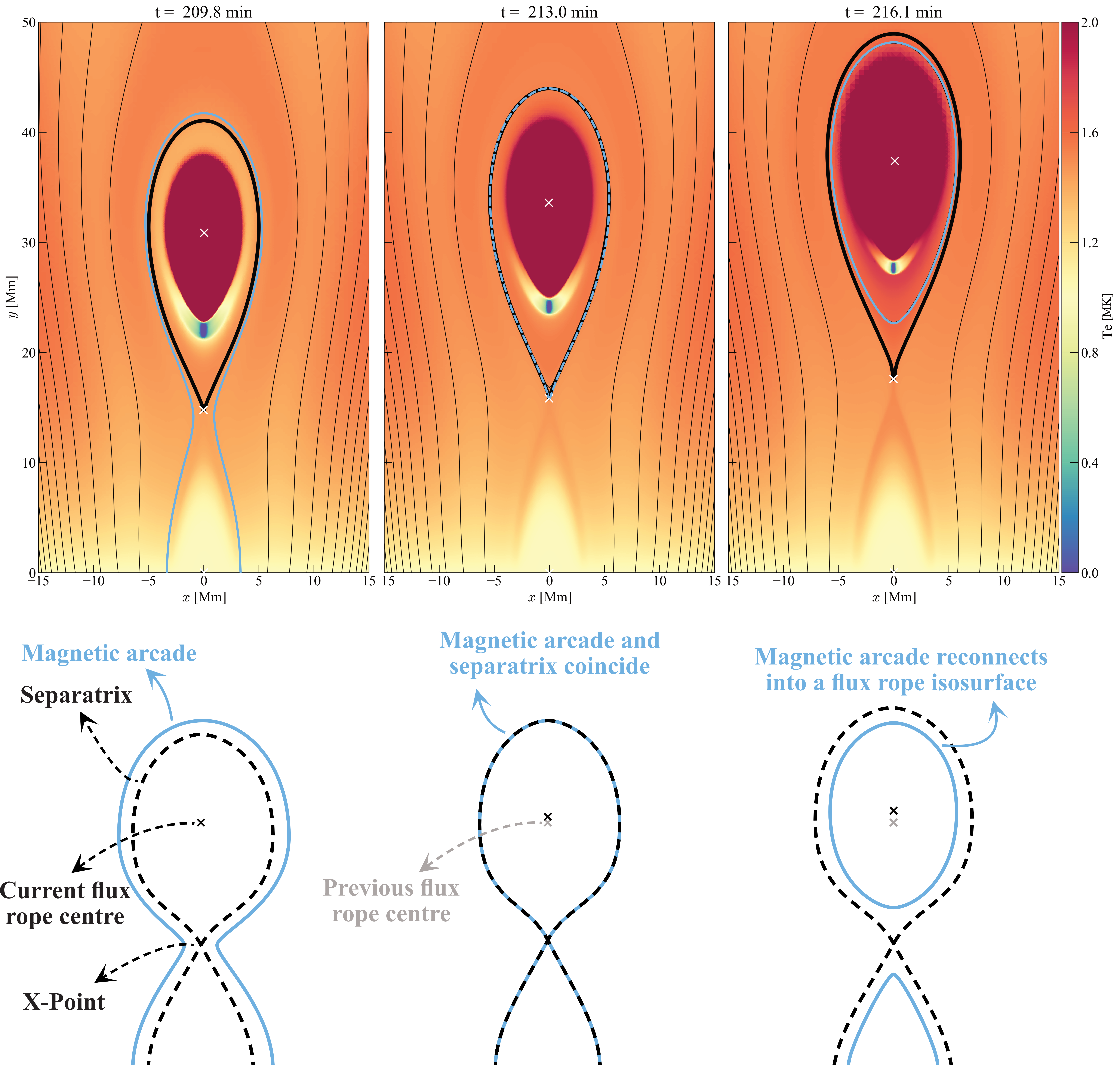}
    \caption{Demonstration of how magnetic reconnection leads to the expansion of the flux rope and consequently how it affects the location of the flux rope centre. Top row shows the temperature evolution in MK for the case $\sigma=1$. Thin black lines are magnetic arcades, the thick black line is the separatrix and the blue magnetic field is a tracked magnetic arcade that eventually reconnects into a closed magnetic field line. The top white cross shows the flux rope centre, defined as the only point  within the flux rope with a vanishing magnetic curvature, and the lower white cross locates the X-point as another, isolated point with zero magnetic curvature. Bottom row is a cartoon that extracts the ongoing process from our simulation and represents it in a simplified manner.}
    \label{fig:slowrisecartoon}
\end{figure}

\begin{figure}[thp]
    \centering
    \includegraphics[width=\linewidth]{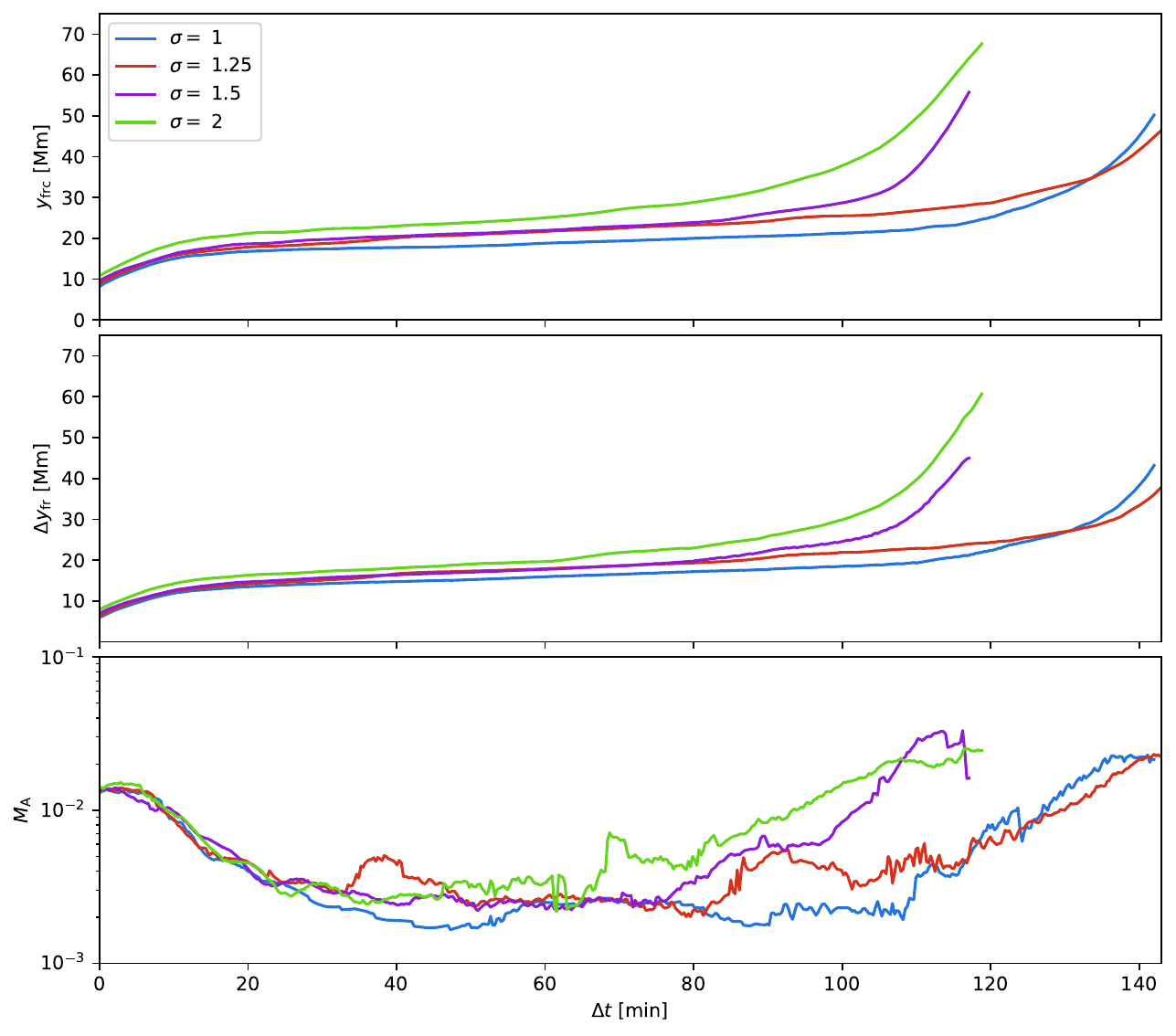}
    \caption{Evolution of the flux rope centre and vertical dimension for the erupting cases $\sigma \in \{ 1, 1.25, 1.5, 2\}$ due to magnetic reconnection. Top panel shows the height evolution of the flux rope centre $y_\mathrm{frc}$, middle panel shows the vertical length of the flux rope $\Delta y_\mathrm{fr}$ and third panel the inflow Alfvén Mach number $M_\mathrm{A}$. The time difference $\Delta t = t - t_\mathrm{fr}$ is with respect to the onset of flux rope formation $t_\mathrm{fr}$. We remark that all curves in the current figure terminate before the formation of their first plasmoids.}
    \label{fig:slowrisealfven}
\end{figure}

{For our work, we are mainly interested in explaining the slow rise phase of the filament. We find that the slow-rise phase is due to the (slow) expansion of the flux rope, which is modulated by magnetic reconnection. In order to demonstrate this, we first need to define the edge of the flux rope. We define it as a separatrix, i.e. the line that demarcates closed magnetic field lines from magnetic arcades. The separatrix is then tracked by first searching for points with zero magnetic curvature, a method inspired by \citet{brughmans2022}. Subsequently, we use this point as a seedpoint\footnote{We have to  displace the seedpoint a tiny bit upwards from the X-point such that we retrieve the flux rope separatrix rather than the closed arcades beneath the X-point. The displacement is equal to the resolution of the finest cell, i.e. 52 km.} for tracing the separatrix that follows the local magnetic field.  Since we have defined the flux rope centre as the only point with vanishing magnetic curvature within the flux rope, it is, then, clear that the location of the flux rope centre is influenced by the size of the flux rope. To exemplify this more specifically, the expansion of the flux rope during the slow rise phase for $\sigma=1$ is demonstrated in Fig. \ref{fig:slowrisecartoon} and complemented with a small cartoon to highlight the ongoing process, though the other erupting cases display the same behaviour. When magnetic arcades magnetically reconnect, they add another magnetic isosurface to the flux rope, resulting in the continual expansion and rise of the flux rope.}
{This explanation on magnetic reconnection modulating the slow-rise phase is further supported through our quantitative study, which can be seen in Fig. \ref{fig:slowrisealfven}. The figure displays the evolution of the flux rope centre's height $y_\mathrm{frc}$ (top panel), the flux rope's vertical length $\Delta y_\mathrm{fr}$ (middle panel) and the inflow Alfvén Mach number $M_\mathrm{A}=v_x / v_\mathrm
{A}$. The vertical length of the flux rope is calculated as the difference between the separatrix's maximum and minimum height, the latter being the X-point. The inflow Alfvén Mach number $M_\mathrm{A}$ is computed by taking a 1D horizontal cut of 2 Mm that intersects the X-point (1 Mm to its left and 1 Mm to its right). By further searching for the maximum\footnote{Searching for the maximum value translates to the largest positive value of $M_\mathrm
{A}$ in the current context. However, our conclusions would still remain valid if the global minimum along the ray would have been used instead due to the global symmetry of our system.} value of $M_\mathrm{A}$ along this ray, we obtain a proxy for the magnetic reconnection rate.} 

{From the figure, it can immediately be seen that magnetic reconnection is the main driver of the height evolution and expansion of the flux rope. Between $\Delta t \in [0, 30]$ we note that magnetic reconnection drops by an order of magnitude from $M_\mathrm{A}\approx 2 \cdot 10^{-2}$ to $M_\mathrm{A}\sim10^{-3}$ from the third panel. Within the same time period, both the flux rope's centre and vertical length decelerate strongly and accordingly for all erupting $\sigma$-cases which can be seen from the first and second panel, respectively. When magnetic reconnection remains low (but non-zero) at $M_\mathrm{A} \sim 10^{-3}$, the flux rope expands and rises slowly and steadily for all shown cases in the same time span. Only when the local Mach number $M_\mathrm{A}$ increases by an order of magnitude again from $M_\mathrm{A}\sim10^{-3}$ to $M_\mathrm{A}\sim10^{-2}$ does the flux rope evolve much more rapidly: its flux rope centre and vertical length increase accordingly with the Mach number. These results further show the smooth transition from the slow-rise phase to the impulsive phase, which is due to the smooth transition from low Mach number $M_\mathrm{A}\sim10^{-3}$ to higher Mach number $M_\mathrm{A}\sim10^{-2}$. In conclusion, we have proven that for our CMEs magnetic reconnection dominates the dynamics of the flux rope: from the slow-rise phase to the impulsive phase.} These results agree with the works of \citet{fan2017, xing2024}, who owed their slow rise due to magnetic reconnection as well.  

From the height-time plot of $\sigma=1.5$ in Fig. \ref{fig:height_evolution}, it shows a clear bump at $\Delta \, t \approx 119 \ \mathrm{min}$. This is even more conspicuous in the vertical-velocity-time plot in the second panel of the same figure. {We have applied a Gaussian smoothing kernel on the velocity curve to smooth out small local gradients.} For the other erupting cases $\sigma \in \{1, 1.25, 2 \}$, instead of the bump a velocity plateau is present at $t=149 \ \mathrm{min}, t=147 \ \mathrm{min}$ and $t=123 \ \mathrm{min}$, respectively. This bump and these velocity plateaus originate from the first plasmoid formation during the impulsive phase of the CME (which can be seen in the second row of Fig. \ref{fig:te_evolution}). When plasmoids are relatively large, it can impede magnetic reconnection \citep{shibata2001} which results in slower driving of the CME in our case. When the CME further evolves, the effects of subsequent plasmoids are less outspoken. Nevertheless, we see that for all the erupting cases we retrieve vertical speeds $v_y$ in the order of $\sim 40 - 80 \ \mathrm{km} \, \mathrm{s}^{-1}$  which is in agreement with observations of erupting prominences \citep{gopalswamy2003}. We further notice that the vertical velocity reaches its maximum at $\Delta \, t\approx 147 \ \mathrm{min}$ for $\sigma \in \{ 1.5, 2\}$ and $\Delta \, t\approx 171 \ \mathrm{min}$ for $\sigma =1$, after which it slightly decreases. This is an indication that the top boundary at $y=300 \ \mathrm{Mm}$ is interfering with the CME evolution and impeding its acceleration. 

\begin{figure}[thp]
    \includegraphics[width=0.235\textwidth]{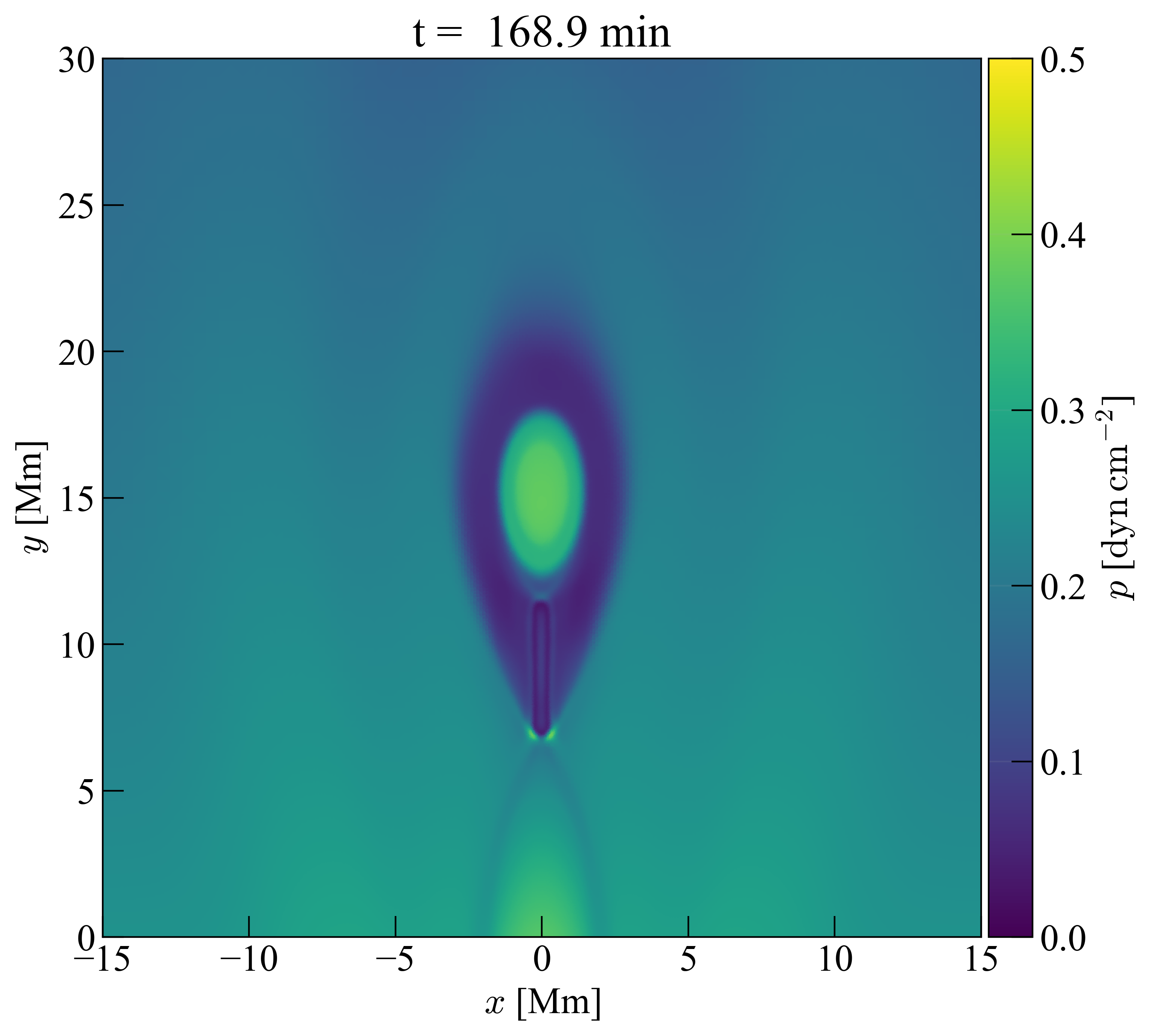}
    \includegraphics[width=0.235\textwidth]{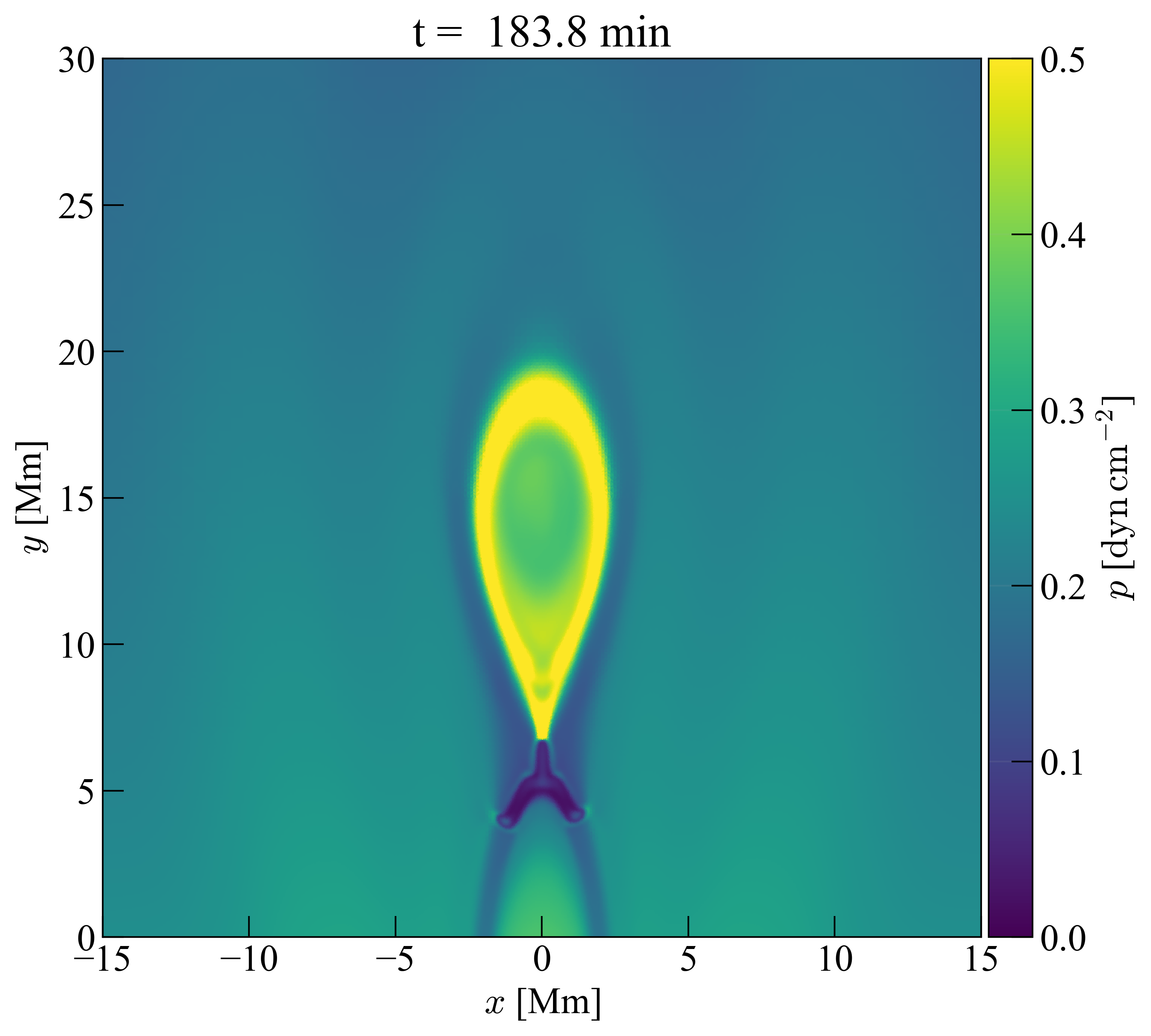} \\
    \includegraphics[width=0.235\textwidth]{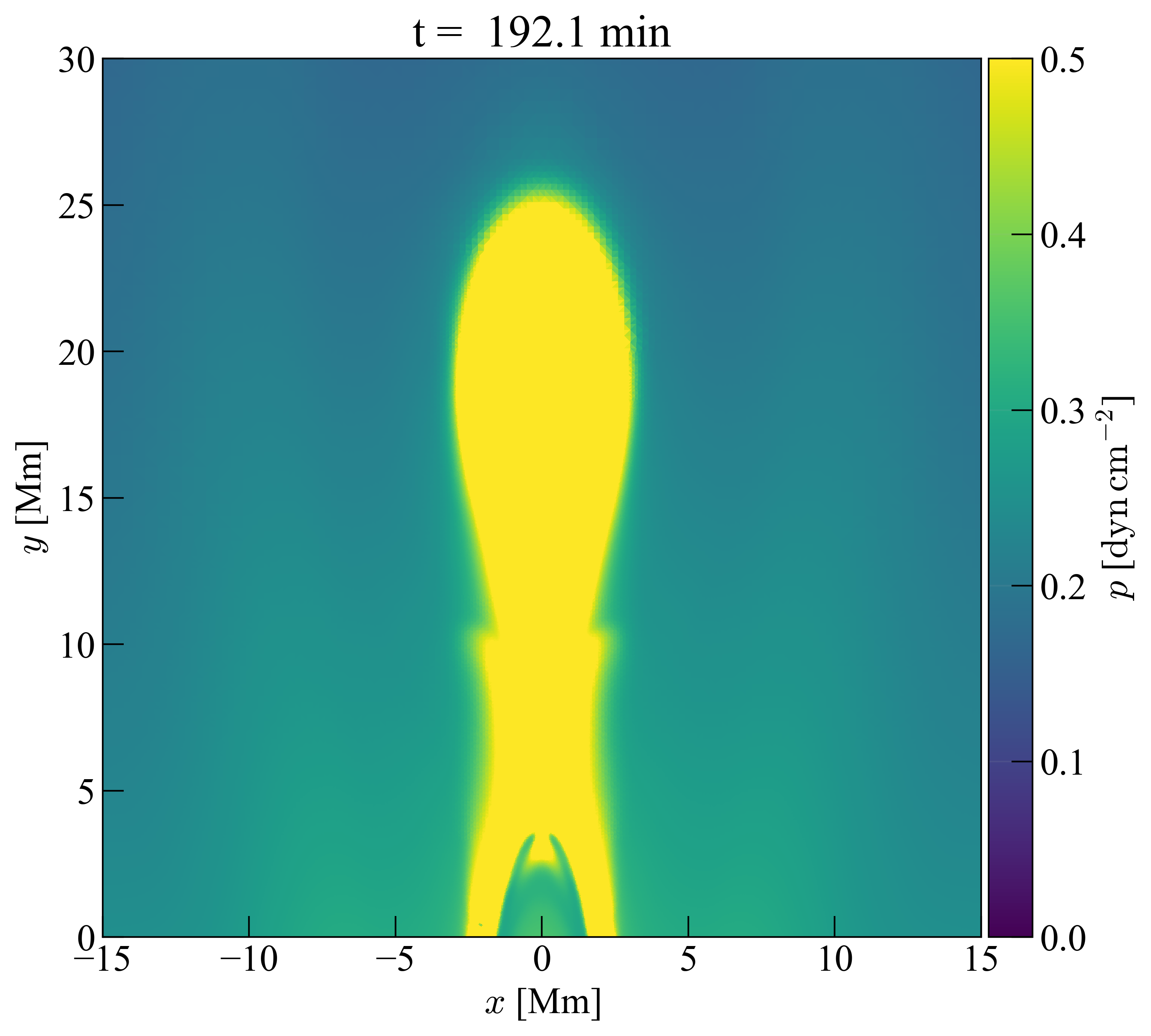}
    \includegraphics[width=0.235\textwidth]{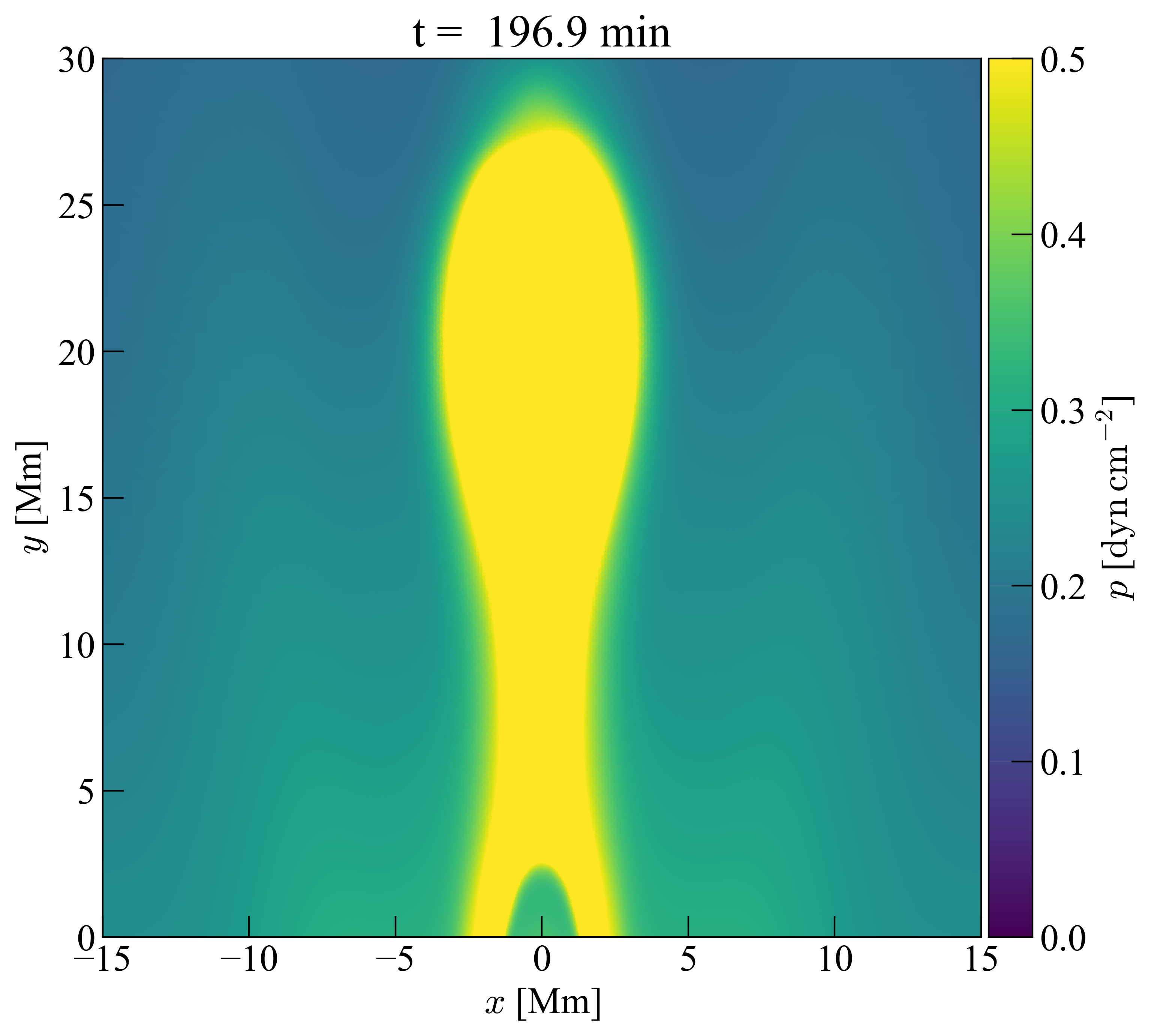}
    \caption{Pressure $p$ colour plots in CGS units at four different timestamps for the $\sigma = 0.75$ case.}\label{fig:pressure_evap}
\end{figure}

{The flux rope centres of $\sigma = 0$ and $\sigma = 0.75$ show an oscillating behaviour, with $\sigma = 0.75$ dampening over time. The cause for the oscillation} is elaborated on in Fig. \ref{fig:pressure_evap} for $\sigma = 0.75$ but the discussion holds for $\sigma=0$. The flux rope first possesses a solar prominence (top left panel) which, as we have mentioned before, escapes the magnetic flux rope (top right panel). When the condensation reaches our lower boundary (where we extrapolate assuming constant 1 MK conditions), it gradually evaporates and substantial amounts of pressure enters the simulation box through the bottom boundary and the bottom boundary further maintains this high pressure state. This in-stream of high gas pressure pushes on the flux rope and brings it out of its equilibrium position (bottom right panel). Therefore, this oscillation is induced by the bottom boundary treatment, here assuming a 1 MK heat reservoir. That the condensation evaporates at the bottom boundary is primarily due to thermal conduction since the bottom boundary adopts an isothermal temperature of $T= 1 \ \mathrm{MK}$, allowing the condensation to gradually heat up. While the recent work of \citet{johnston2025} showed that filaments can escape their magnetic topology by driving underlying magnetic reconnection, in our case it is more likely that the filament escapes due to mass slippage which has been obtained in other works that assume almost the same magnetic topology as we do \citep{jenkins2021, jenkins2022, donne2024}.   However, it is unable to dominate over the overlying strapping field and therefore is pushed back downwards, creating this oscillatory behaviour. {The oscillation of the flux rope of $\sigma=0.75$ dampens over the time due to the gradual reduction of pressure effects from the bottom boundary whereas the bottom boundary of $\sigma=0$ maintains inflow of energy, which drives the oscillation of the flux rope. }

Overall, our results based on its dynamics agree very well with observed erupting prominences. One specific observed erupting prominence is shown in the bottom panel of the same figure, with courtesy of \citet{DiLorenzo2025}.  There, it can be seen that {from the datapoint $\Delta t \approx 1 \, 094 \ \mathrm{min}$ onwards the prominence exhibits a height-time profile that is conspicuously non-linear. Hence, we roughly estimate that the slow rise phase of this observed case lasts for $\approx 1 \, 094 \ \mathrm{min}$}. Despite our idealised 2.5D configuration (with a flux rope that is disconnected from the surface) and absence of any fine-tuning, the timescale of our prominence eruption $\tau \sim 180 \ \mathrm{min}$ is much faster than this specific observation, but shows a similar trend. It will be of interest to extend this work to a more realistic 3D model as line-tying effects will delay eruptions (as well as allow prominence drainage along the flux rope axis, see work by \citet{fan2017,fan2018}). 

\subsubsection*{Energetics of the CMEs}

\begin{figure}[thp]
    \includegraphics[width=0.48\textwidth]{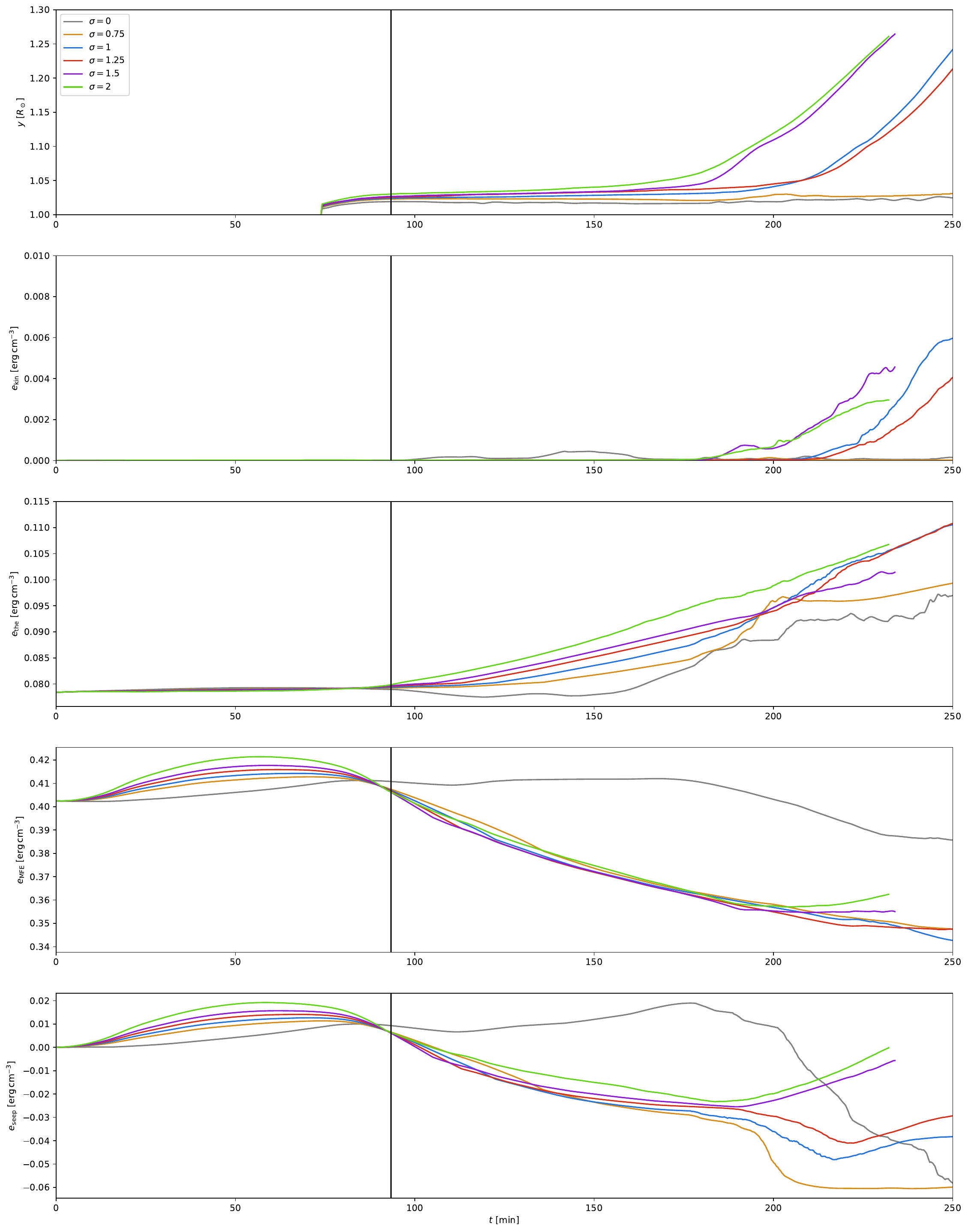}
    \caption{Height evolution (top panel),  average kinetic energy density (second panel), average thermal energy density (third panel), {average magnetic free energy density (fourth panel) and average seeped energy density (last panel)} for the six different $\sigma$-cases. The vertical black line indicates the instant when footpoint motion has been turned off at $t=t_1$. Energy densities are expressed in erg cm$^{-3}$ and are averaged over the entire simulation domain $\vert x \vert <30 \ \mathrm{Mm}$ and $y<300 \ \mathrm
    {Mm}$.}\label{fig:energies}
\end{figure}

We can further understand the height-time plot if we compute the {magnetic free-energy} $e_\mathrm
{MFR}$, kinetic- $e_\mathrm{kin}$ and thermal energy densities $e_\mathrm{the}$ within the entire simulation box, i.e. $\vert x \vert \leqslant 30 \ \mathrm{Mm}$ and $\vert y \vert \leqslant 300 \ \mathrm{Mm}$, and the average energy density seepage $e_\mathrm{seep}$ that quantifies how much energy has already flowed out of our domain on average. {The magnetic free energy is computed by first constructing the potential field as outlined by \citet{chiu1976,xia2018} and subsequently subtracting the total magnetic energy from the potential energy. The average energy density seepage considers all flux terms in the energy equation, i.e. compressional heating, Poynting flux, thermal conduction, Ohmic heating, background heating, radiative cooling and gravitational work. All these flux contributions are summed into a net flux and subsequently integrated over the entire simulation domain to obtain the average energy seepage rate. However, since the other quantities are energy densities, we must express the amount of seepage in the same units. We do so by dividing the average energy seepage rate by the simulation volume and integrating it with time. This gives us the average energy density seepage $e_\mathrm{seep}$ which is a cumulative quantity. In mathematical form, the energy seepage is}
\begin{align}
    e_\mathrm{seep}(t) &\equiv V^{-1}\int_0^t \int_\mathcal{V} \bigg(-(\nabla \cdot \mathbf{F})\vert_\mathrm{net} + S_\mathrm{net}\bigg) \,\mathrm{d}V \mathrm{d}t
\end{align}
{with $\mathcal{V}$ the simulation domain, $(\nabla \cdot \mathbf{F})\vert_\mathrm{net}$ the net flux from all the conservative terms and $S_\mathrm{net}$ the net flux from all the source- and sink terms. In the current definition, $e_\mathrm{seep}(t) > 0$ indicates energy seeping into the domain and, conversely, $e_\mathrm{seep}(t) < 0$ specifies energy leaking out. Note that we numerically evaluate the time integral in $e_\mathrm{seep}$ by integrating over the (uniform) times between the output files rather than the CLF timestep of the simulation. Hence, the cumulative energy seepage density is a rough estimation.} The results are  provided by Fig. \ref{fig:energies}, complemented with the same height-time plot as in the top panel of Fig. \ref{fig:height_evolution}. The flux rope forms roughly after $t\approx 73 \ \mathrm{min}$ and thus that is when the tracking starts. For $t \lesssim 73 \ \mathrm{min}$, the flux rope is absent, so its centre is set to the bottom of our simulation domain, i.e. $y=0$ or $y= 1 \ \mathrm{R}_\odot$. 

The second panel exhibits the kinetic energy density of the simulation. For the erupting cases $\sigma \in \{1, 1.25, 1.5,2\}$, the kinetic energy density starts to become significant once the CME is going through its impulsive phase and reaches energies in the order of $e_\mathrm{kin} \sim 10^{-3} \ \mathrm{erg} \, \mathrm{cm}^{-3}$. For the grey ($\sigma=0$) curve, there are some early fluctuations during $98 \ \mathrm{min} \lesssim t \lesssim 167 \ \mathrm{min}$. These originate from coronal rain which have a similar density as the solar prominence but reach significant speeds with the fastest ones around $\vert \mathbf{v} \vert \sim 60 \ \mathrm{km} \, \mathrm{s}^{-1}$. They are formed at around $t \approx  \ 98 \ \mathrm{min}$ and fall to the bottom part of our simulation domain at around $t \approx  \ 167 \ \mathrm{min}$. Almost no noticeable kinetic energy is detected in the beige ($\sigma=0.75$) curve. 

Notice that the erupting case where the prominence has evaporated ($\sigma = 1$) has a higher final kinetic energy density than the other erupting cases where the prominence did not evaporate ($\sigma \in\{1.5,2 \}$). While the total kinetic energy density is computed for the entire simulation domain, the erupting flux rope is the dominant component. This higher kinetic energy comes from the fact that the erupting, evaporated prominences have higher peak speeds ($80 - 85 \ \mathrm{km} \, \mathrm{s}^{-1}$) than the erupting, non-evaporated prominences ($\approx 65 \ \mathrm{km} \, \mathrm{s}^{-1}$), which can be seen from Fig. \ref{fig:height_evolution}. {However, the final kinetic energy of $\sigma=1.25$ is lower than $\sigma=1$. This is due to the fact that the prominence of $\sigma=1$ case evaporates before the prominence of $\sigma=1.25$ does, which can also be seen in the provided animation of Fig. \ref{fig:te_evolution}. } Therefore, our results show that the prominence inhibits the flux rope from attaining a fast acceleration. These results agree with \citet{zhao2017}, who showed that the solar prominence dominates the evolution of the flux rope.

The third panel displays the thermal energy density. During footpoint driving (before the black vertical line) there is minor variation between the different $\sigma$-cases. However, after the flux rope has formed the curves start to diverge from each other significantly. The grey ($\sigma=0$) curve is the only curve that reaches a thermal energy density below the initial thermal value ($e_\mathrm{the, 0}$) between $98 \ \mathrm{min} \lesssim t \lesssim 167 \ \mathrm{min}$, which is due to the formation of coronal rain. After $ 167 \ \mathrm{min} \lesssim t$ the thermal energy density increases, despite that no (considerable) magnetic reconnection occurs. This originates partly from the pressure-driven effects that we discussed in Fig. \ref{fig:pressure_evap} for $\sigma=0.75$ and partly from our open side boundaries\footnote{Although the ghost cells at the side boundaries have a velocity $v_x = 0$, there will still be a non-zero flux passing through the boundaries $x=\pm 30 \ \mathrm{Mm}$, as cell center and cell edge values may differ.} after $t_1 < t$. As a result the thermal energies increase for both cases. Note that the thermal energy density of the grey curve flattens, whereas the beige ($\sigma=0.75$) curve steadily keeps increasing. This difference arises from the fact that magnetic reconnection is still going on for $\sigma=0.75$:  its thermal energy increases in a similar fashion to the other erupting cases ($\sigma\in\{1, 1.25, 1.5, 2\}$) and because for these five $\sigma$-cases the magnetic {free} energy density ({fourth panel, see below}) is continuously decreasing, converting its kinetic and thermal energy density, which is not the case for $\sigma=0$. Since the flux rope of $\sigma=0$ undergoes mass depletion but does not erupt, this could also be related to the flux rope attaining a height in which mass depletion does not bring it out of equilibrium as shown by the work of \citet{jenkins2019}. They argue that a critical height exists {above} which mass depletion can lead the flux rope to erupt. Since our case of $\sigma=0$ does not erupt, it is clear that it has not reached this critical height. 
In the work of \citet{fan2018}, prominence draining allowed for the onset of the kink instability. 
{In addition, prominence draining can also facilitate the onset of the torus instability by reducing the stabilizing effect of gravity, as demonstrated in recent 3D simulations by \citet{xing2025}.} 
Further study is needed to extend this work into 3D and examine whether our flux rope tends to undergo either kink {or torus }instability.

While the kinetic- and thermal energy density only start to become noticeable after $t=t_1$, only the {magnetic free energy density} $e_\mathrm{MFE}$ displays changes amongst the different $\sigma$-cases during footpoint driving $t \leqslant t_1$. It can directly be seen that, during the footpoint driving $t \leqslant t_1$, the higher the shearing $\sigma$, the higher the magnetic {free} energy density $e_\mathrm{MFE}$ and hence the higher the energy state of the flux rope as expected. When the footpoint driving is disabled ($t_1 < t$), all cases $\sigma$ except for $\sigma=0$ steadily decrease. Indeed, the theory of CMEs requires the magnetic energy to decrease as the flux rope searches for a new equilibrium at higher altitudes. And as mentioned before, even the beige ($\sigma=0.75$) curve has a decreasing magnetic {free} energy density, despite that it did not erupt. Therefore, we expect that if the simulation time were to be extended, it should erupt eventually. This hypothesis is also motivated by the slight rise of its flux rope centre, whose altitude has increased from $16 \ \mathrm{Mm}$ at the disabling of the footpoint motion to $20 \ \mathrm{Mm}$ by the end of its simulation. The grey ($\sigma=0$) curve shows a decrease in magnetic {free} energy density around $t\approx 175 \ \mathrm{min}$ but after $t\approx 225 \ \mathrm{min}$ it remains quasi-stable again. And since the flux rope centre of $\sigma=0$ does not show considerably increase in altitude, we believe that it will not erupt at all unless footpoint shearing were to continue indefinitely. 

The last panel shows the average, cumulative energy density seepage $e_\mathrm{seep}$. During footpoint driving (before the black vertical line), we find energy flowing into the domain from our bottom boundary treatment, inducing shearing of the magnetic field lines as expected. Once footpoint driving has been disabled, all cases, except for $\sigma=0$, show negative seepage values, indicating energy flowing out of our domain.

From these results, we can quantify the conversion rates from magnetic {free} energy density to thermal and kinetic energy density and examine the efficiency of magnetic flux cancellation in driving a CME. To obtain a consistent result, we define the change in magnetic {free} energy density $\Delta \, e_\mathrm{MFE}$ as the difference between the global minimum and maximum of the curve, i.e. $\Delta \, e_\mathrm{MFE} \equiv e_{\mathrm{MFE, min}} - e_{\mathrm{MFE, max}}$, with $e_{\mathrm{MFE, max}}$ and $e_{\mathrm{MFE, min}}$ the maximal and minimal attained magnetic {free} energy density, respectively. Their corresponding timestamps $\tau_{\mathrm{MFE, max}}$ and $\tau_{\mathrm{MFE, min}}$ are then used to find the instantaneous thermal and kinetic energy densities and define the change in energy densities $\Delta \, e_\mathrm{the}$ and $\Delta \, e_\mathrm{kin}$, respectively. By further defining the energy conversion percentage $\epsilon$ as 
\begin{equation}
    \epsilon \equiv 100 \cdot \bigg\vert \dfrac{(\Delta e / \Delta \, \tau)\vert_\mathrm{kin/the}}{(\Delta e / \Delta \, \tau)\vert_\mathrm{MFE}} \bigg\vert
\end{equation}
we can quantify the conversion rates. These quantifications can be observed in Table \ref{table:energy_conversion}. {Since our boundaries are partially open, the energy seepage density $e_\mathrm{seep}$ from Fig. \ref{fig:energies} has shown that  our simulation has lost some energy: either through the contribution from sink terms such as radiative cooling or energy flowing out through the boundaries.} Therefore,our rates from Table \ref{table:energy_conversion} are {merely} indicative.
\begin{table*}[tbp]
 \caption{Energy conversions from magnetic to kinetic and thermal energy densities in function of the shearing $\sigma$.} 
 \begin{center} \footnotesize
    \begin{tabular}{P{0.03\textwidth} P{0.04\textwidth} P{0.04\textwidth} P{0.07\textwidth} P{0.07\textwidth} P{0.07\textwidth} 
                    P{0.08\textwidth} P{0.08\textwidth} P{0.04\textwidth} P{0.07\textwidth} P{0.07\textwidth} P{0.04\textwidth}}
        \toprule 
        \toprule
         & & & \multicolumn{3}{c}{Magnetic} & \multicolumn{3}{c}{Kinetic} & \multicolumn{3}{c}{Thermal} \\
        \cmidrule(lr){4-6} \cmidrule(lr){7-9} \cmidrule(lr){10-12}
        $\sigma$ & $\tau_{1}$ & $\Delta \, \tau$ & $e_{\mathrm{MFE, 1}}$ & $\Delta \, e$ & $\Delta \, e / \Delta \, \tau$ & $\Delta \, e$ & $\Delta \, e / \Delta \, \tau$  &  $\epsilon$ & $\Delta \, e$ & $\Delta \, e / \Delta \, \tau$ & $\epsilon$  \\
        \midrule
$0   $ & $167.7$ & $ 85.6$ & $4.12 \cdot 10^{-1}$ & $-2.7 \cdot 10^{-2}$ & $-5.2 \cdot 10^{-6}$ & $7.3 \cdot 10^{-5}$ & $1.4\cdot 10^{-8}$ & $0.27$ & $1.5 \cdot 10^{-2}$ & $3.0 \cdot 10^{-6}$ & $57.5$ \\[3mm]
$0.75$ & $ 71.0$ & $182.3$ & $4.13 \cdot 10^{-1}$ & $-6.5 \cdot 10^{-2}$ & $-6.0 \cdot 10^{-6}$ & $1.4 \cdot 10^{-5}$ & $1.3\cdot 10^{-9}$ & $0.02$ & $2.1 \cdot 10^{-2}$ & $1.9 \cdot 10^{-6}$ & $31.6$ \\[3mm]
$1   $ & $ 67.6$ & $185.8$ & $4.14 \cdot 10^{-1}$ & $-7.3 \cdot 10^{-2}$ & $-6.5 \cdot 10^{-6}$ & $5.9 \cdot 10^{-3}$ & $5.3\cdot 10^{-7}$ & $8.16$ & $3.2 \cdot 10^{-2}$ & $2.9 \cdot 10^{-6}$ & $44.4$ \\[3mm]
$1.25$ & $ 61.8$ & $185.8$ & $4.16 \cdot 10^{-1}$ & $-6.9 \cdot 10^{-2}$ & $-6.2 \cdot 10^{-6}$ & $3.7 \cdot 10^{-3}$ & $3.3\cdot 10^{-7}$ & $5.35$ & $3.1 \cdot 10^{-2}$ & $2.8 \cdot 10^{-6}$ & $45.7$ \\[3mm]
$1.5 $ & $ 59.0$ & $150.9$ & $4.18 \cdot 10^{-1}$ & $-6.3 \cdot 10^{-2}$ & $-7.0 \cdot 10^{-6}$ & $1.5 \cdot 10^{-3}$ & $1.7\cdot 10^{-7}$ & $2.45$ & $1.9 \cdot 10^{-2}$ & $2.1 \cdot 10^{-6}$ & $29.5$ \\[3mm]
$2   $ & $ 56.7$ & $148.8$ & $4.21 \cdot 10^{-1}$ & $-6.4 \cdot 10^{-2}$ & $-7.2 \cdot 10^{-6}$ & $1.0 \cdot 10^{-3}$ & $1.1\cdot 10^{-7}$ & $1.56$ & $2.1 \cdot 10^{-2}$ & $2.4 \cdot 10^{-6}$ & $33.3$ \\[3mm]
        \bottomrule
        \bottomrule
\end{tabular}
\tablefoot{$e_\mathrm{MFE, 1}$ is the attained maximum {free} magnetic energy density. The timestamp $\tau_{\mathrm{MFE, max}}$ indicates the corresponding moment when the magnetic {free} energy density has reached its maximum. $\Delta \, \tau$ signifies the time period between the maximum and minimum of the magnetic {free}  energy density. $\Delta \, e$ refers to the change in magnetic/kinetic/thermal energy density at times when the magnetic  {free} energy density is maximal and minimal. These quantities are then used to calculate the average energy conversion rate $\Delta e / \Delta \, \tau$ for each energy type. For the kinetic and thermal energy density, $\epsilon$ quantifies the percentage of energy conversion rate from magnetic to kinetic/thermal. Time $\tau$ and time periods $\Delta \, \tau$ are expressed in minutes, energy density $e$ and energy density differences $\Delta \, e$  in ergs cm$^{-3}$, energy density rates $\Delta e / \Delta \, \tau$ in erg cm$^{-3}$ s$^{-1}$ and the energy conversion percentage $\epsilon$.}\label{table:energy_conversion}
\end{center}
\end{table*}

From the table, it is clear that increased shearing $\sigma$ results in a higher obtained maximum magnetic {free} energy density $e_{\mathrm{MFE, max}}$ at shorter times $\tau_{\mathrm{MFE, max}}$. However, $e_{\mathrm{MFE, max}}$ varies fairly little amongst the different cases $\sigma$ with the largest relative difference being $2.1 \%$. It is important to emphasize that this quantity relates to the total energy of the magnetic field. Nevertheless, this small, relative difference in total magnetic energy gives rise to a rich variety of phenomena, i.e. eruption versus no eruption, prominence evaporation versus prominence maintenance and morphological differences as we will show later. 
All erupting cases {exhibit a systematically larger decrease in magnetic free energy density than the non-erupting cases, suggesting an energetic separation between erupting and non-erupting regimes}.

To compare energy conversions among the different $\sigma$-cases, using $\Delta \, e / \Delta \, \tau$ is more relevant as it normalizes the change in energy density with respect to the corresponding time period. In particular, $\epsilon$ makes comparison more simple and convenient as it is expressed in percentage. There, it can immediately be seen that the energy conversion from magnetic to kinetic energy density is small: up to $\approx 10\%$ of the magnetic {free} energy density rate. Instead, a large fraction of the magnetic {free} energy density for the erupting cases is {partly} converted into thermal energy density heating up the plasma with values ranging from $29\% - {57}\%$, {and partly lost by energy seeping out of our domain as shown in Fig. \ref{fig:energies}. Moreover, our relatively high resistivity results in higher magnetic diffusion.} Therefore, {the inefficiency of converting magnetic free energy into kinetic energy to push the flux rope could be particular to our simulation setup. Different boundary conditions and resistivity values from our present study can influence this efficiency by limiting energy seeping out of the domain and constraining magnetic energy conversion into heat, respectively.} Furthermore, it should be noted that an increasing $\sigma$ does not imply a proportionate increase in kinetic and thermal energy density. This can explicitly be seen from $\Delta \, e / \Delta \tau$ and $\epsilon$ for the kinetic and thermal energy densities: there is no monotonic increase. Lastly, we note that the range of our magnetic {free} energy density $e \sim 0.4 \ \mathrm{erg} \, \mathrm{cm}^{-3}$ is in agreement with observed CME energy densities \citep{chen2011}.

\begin{figure}[thp]
    \includegraphics[width=0.24\textwidth]{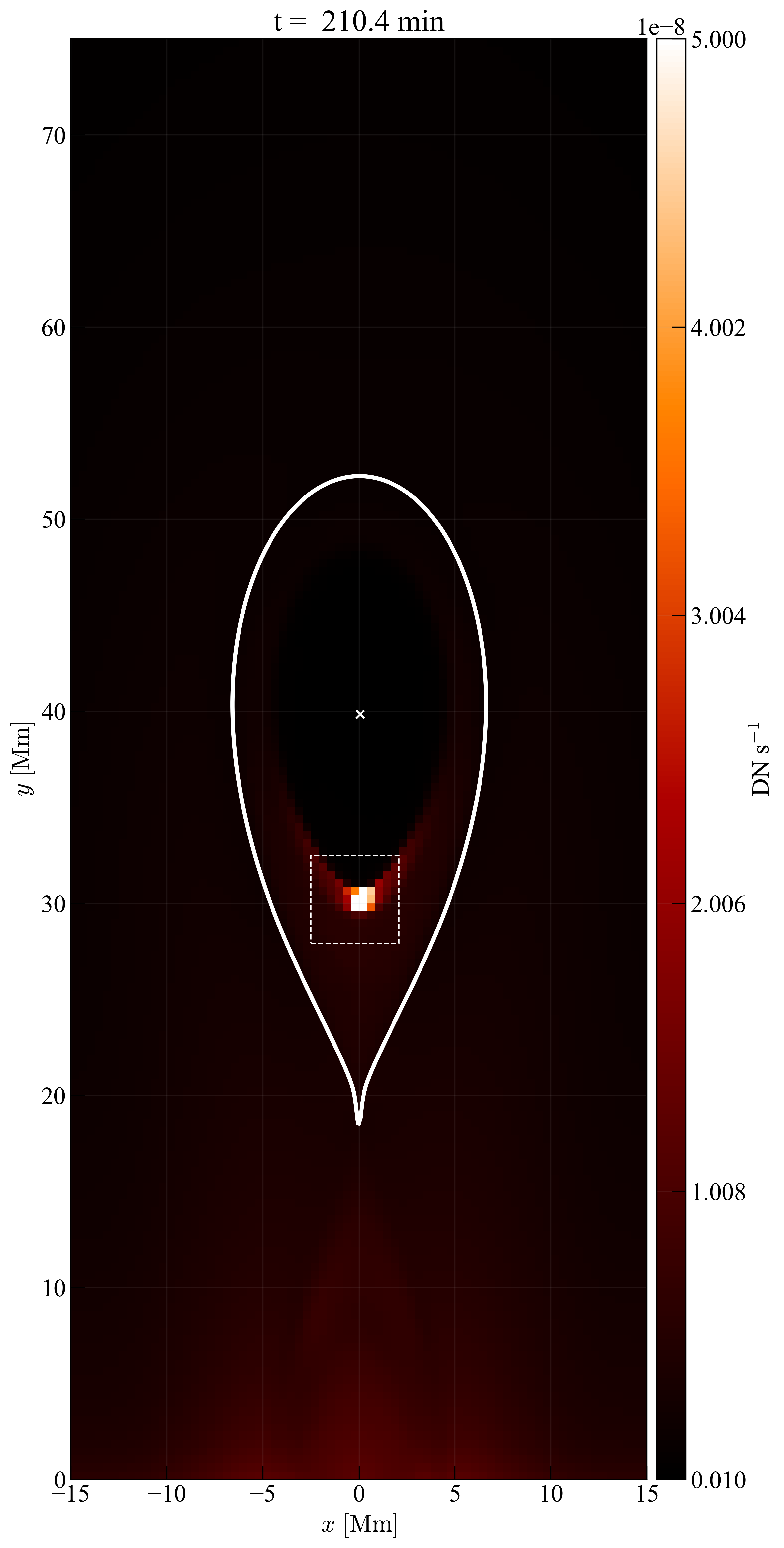}
    \includegraphics[width=0.24\textwidth]{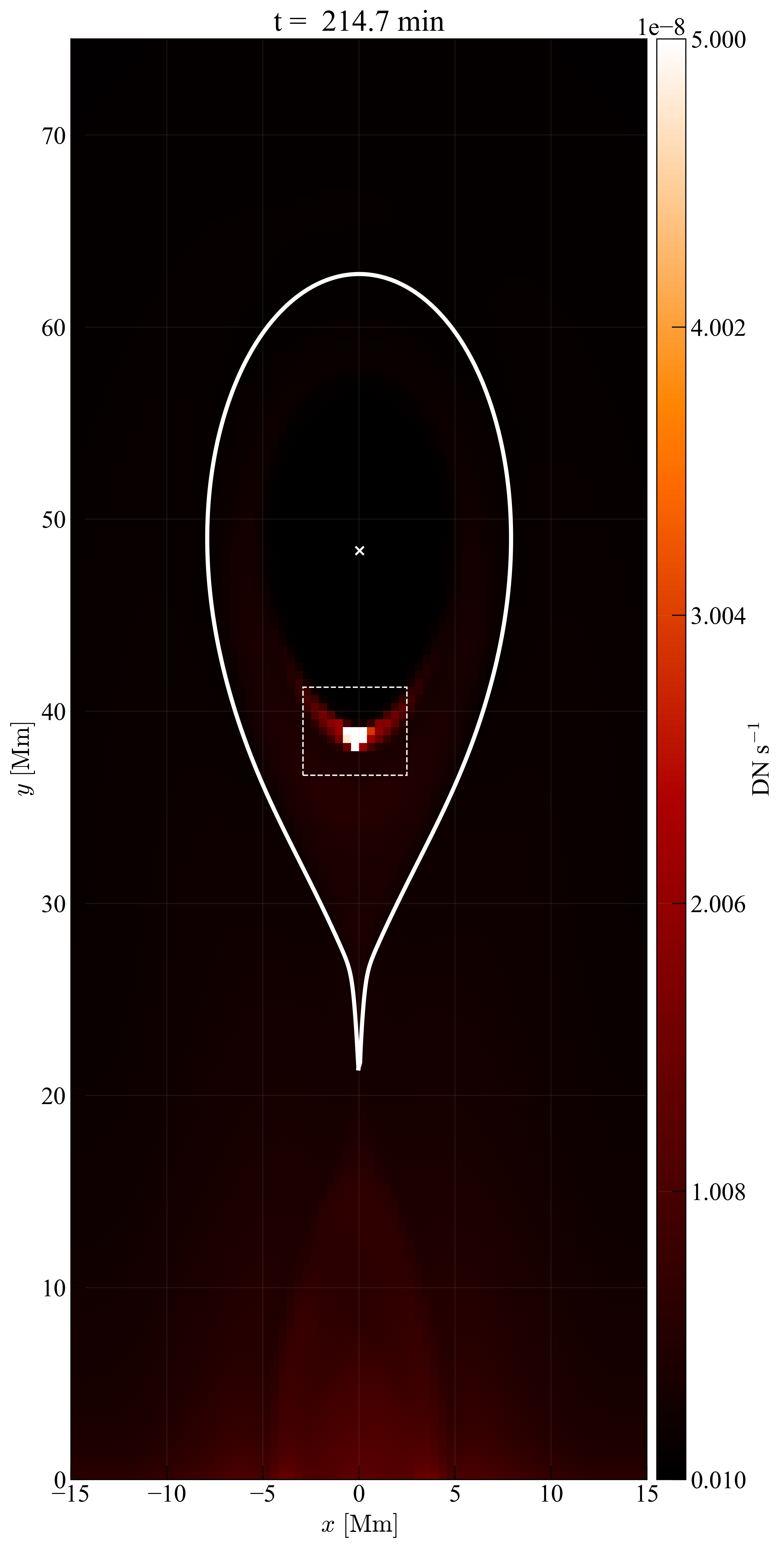} 
    \includegraphics[width=0.24\textwidth]{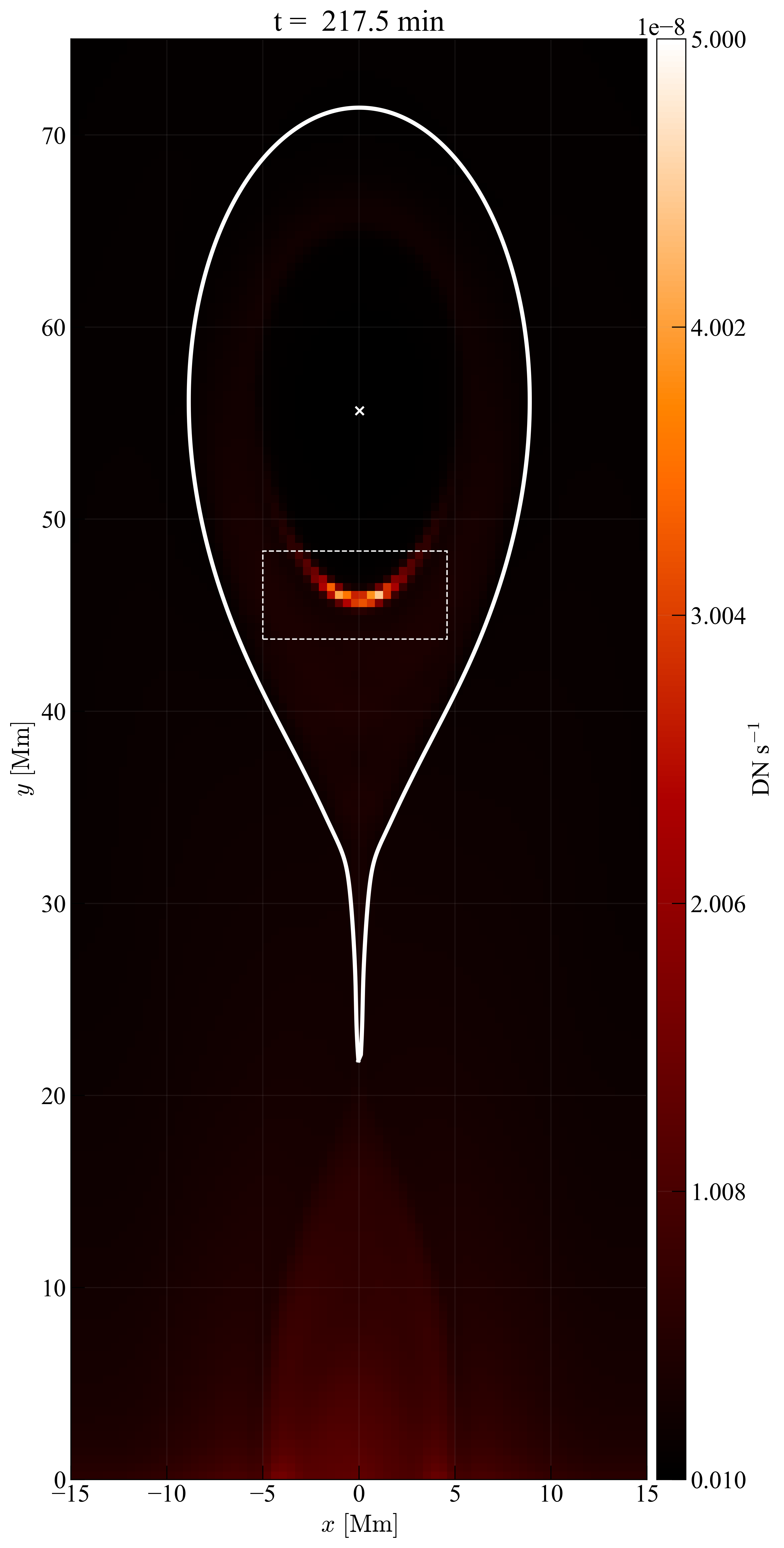}
    \includegraphics[width=0.24\textwidth]{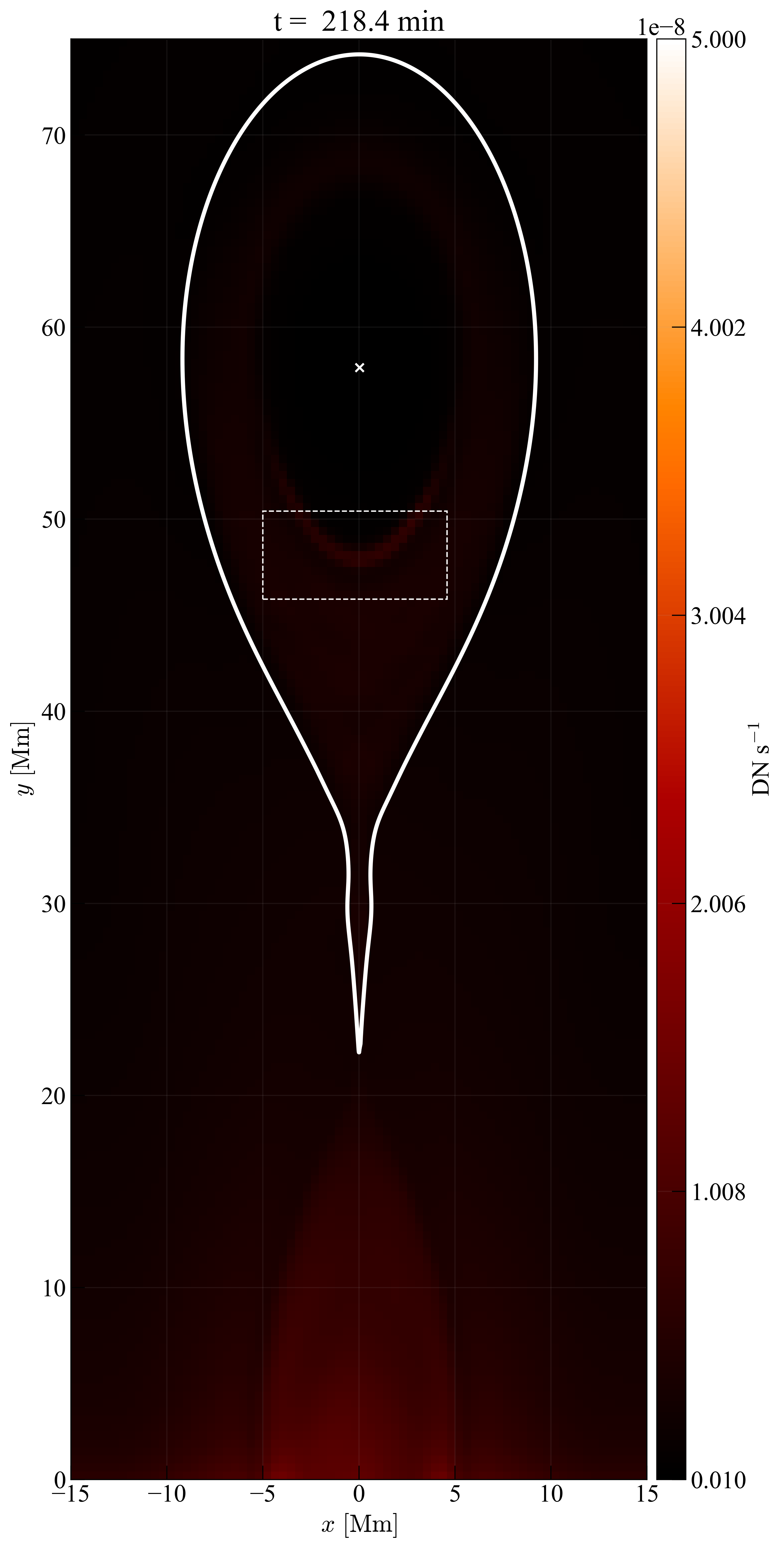}
    \caption{Tracking of the prominence for $\sigma=1$ as shown in the synthetic 304 \ \AA \ channel. The white, dashed rectangle displays the region over which we average in Fig. \ref{fig:evaporation}. The $x$ and $y$ coordinates are shown in megameters. Solid white line shows the separatrix of the magnetic flux rope.} \label{fig:track_rectangle}
\end{figure}

\subsection{Varying thermodynamical environments of flux rope interiors}

Interestingly, only for $\sigma = 0$ does coronal rain occur but in all other cases $\sigma > 0$ coronal rain does not develop. Therefore, from our results it seems that coronal rain formation is only promoted at low energy states $0 \leqslant \sigma < 0.75$, which is likely coupled to the ambient heating resulting from flux rope formation. As \citet{low2001} explains, the interior of the flux rope is in thermal isolation due to long magnetic field lines and due to inefficiency of thermal conduction perpendicular to the magnetic field. He further argues that within this thermally isolated environment instabilities relating to condensations are suitable to occur such as solar prominences. Coronal rain, on the other hand, forms outside the flux rope and hence is not thermally isolated. The reduced dimensionality of our setup does also play a role: our previous 3D simulation \citep{donne2024} did show evidence for rain forming in locations above the forming flux rope as pressure-induced thermodynamic changes led to thermal instability onset. Future 3D work with varying $\sigma$ driving effects must reconsider how prone flux rope formation is to rain formation.

For the higher shearing cases $\sigma=1$ and $\sigma = 1.25$ the prominences evaporate during the initiation of the eruption. In the literature a solar prominence has been observed to evaporate \citep{wang2016}, though the authors could only speculate about the underlying reasons, one of their reasons being thermal conduction between the hot corona and cold prominence. {From our simulated evaporated prominences, our results show that thermal conduction and compressional heating are at play.}

\begin{figure}[thp]
    \includegraphics[width=0.48\textwidth]{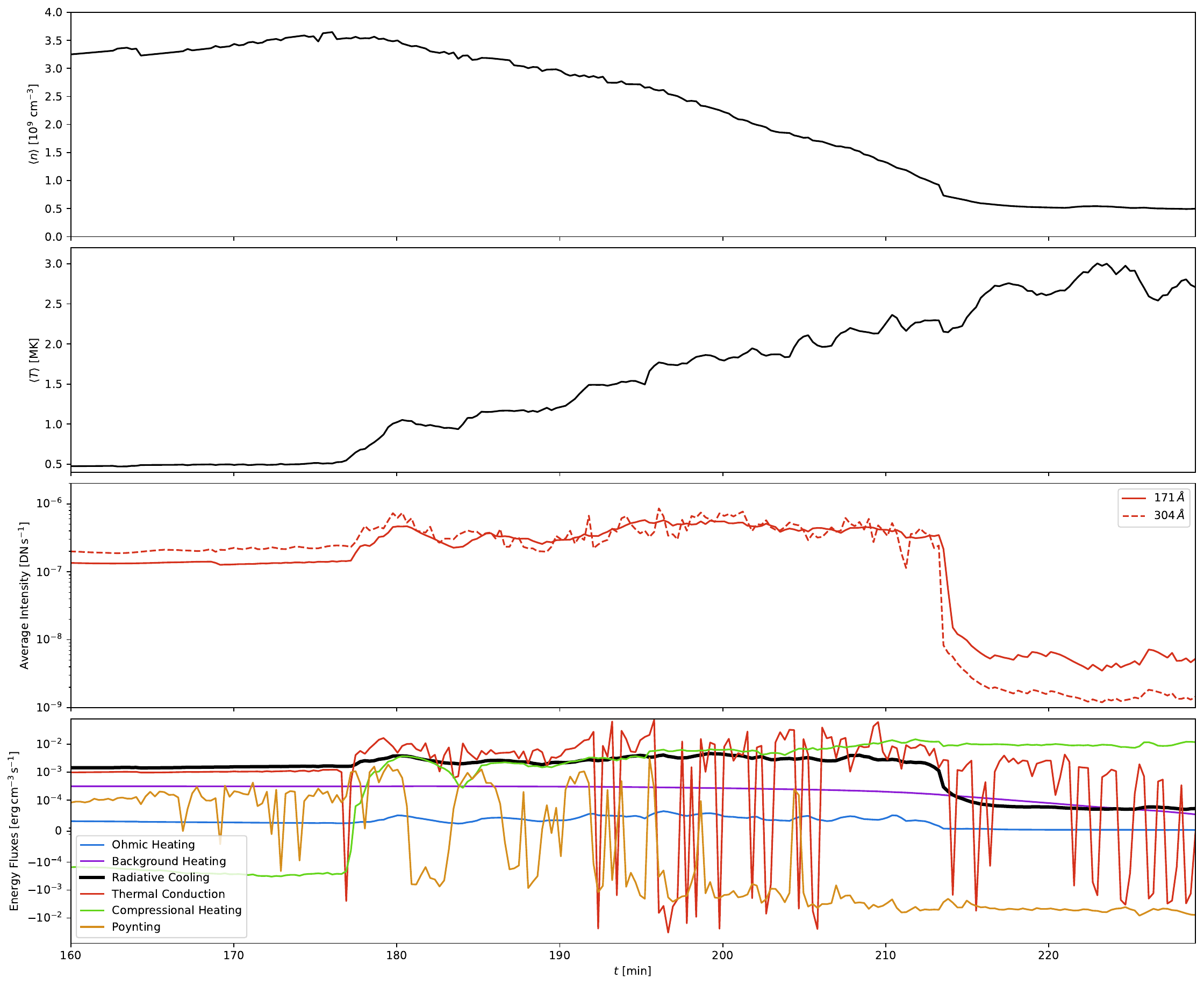}
    \caption{Evolution of the average density in $10^9$ cm$^{-3}$ (first panel), average temperature in MK (second panel), average synthetic 171 \AA\ and 304 \AA\ intensities in DN s$^{-1}$ (third panel), {and the relevant energy fluxes from the energy equation.} All values are averaged within a region around the prominence as shown in Fig. \ref{fig:track_rectangle}. {Note that in the last panel we display the absolute value of radiative cooling.} The data includes averaged values before, during and after the evaporation.}\label{fig:evaporation}
\end{figure}

To prove that we simulate prominence evaporation, we study the case  $\sigma=1$ and define a rectangle around the evaporation as shown in Fig. \ref{fig:track_rectangle} where we display the synthetic 304 \AA \ channel, one of the SDO/AIA channels that was used to examine the prominence evaporation in \citet{wang2016}. Within this rectangle, we average over the number density $n$, temperature $T$, the 171 \AA \ and 304 \AA \ synthetic intensities {and lastly all the energy fluxes.} The results can be seen in Fig. \ref{fig:evaporation}.

The average number density $\langle n \rangle$ and temperature $\langle T \rangle$ can be seen in the first- and second panel, respectively, and help us demarcate the stages of evaporation. For $t<176 \ \mathrm{min}$, the average density {and temperature} maintain a quasi-stable value of $\langle n \rangle \approx 2\,\cdot 10^{9} \ \mathrm{cm}^{-3}$ and $\langle T \rangle \approx 0.5 \ \mathrm{MK}$. Between $176 \ \mathrm{min} \leqslant t \leqslant 219 \  \mathrm{min}$, the density steadily decreases whereas the temperature increases, a trend we would expect for an evaporating fluid. After $t\geqslant 219 \ \mathrm{min}$, all the prominence material has vanished, resulting in a typical coronal density of $\sim 10^8 \ \mathrm{cm}^{-3}$ and temperature of $\sim 2 \ \mathrm{MK}$. This shows that our simulations self-consistently retrieve an evaporating prominence. The time $t<176 \ \mathrm{min}$ marks the period before evaporation, $176 \ \mathrm{min} \leqslant t \leqslant 219 \ \mathrm{min}$ during evaporation and $t\geqslant 219 \ \mathrm{min}$ after evaporation.

The third panel in the same figure shows the averaged synthetic intensities. Before the evaporation of the prominence, the averaged synthetic intensities assume their baseline values around $10^{-7} \ \mathrm{DN} \, \mathrm{s}^{-1}$. Between $176 \ \mathrm{min} \leqslant t \leqslant 219 \ \mathrm{min}$ the prominence is evaporating and the intensities increase significantly for a duration of $\approx 43 \ \mathrm{min}$. Once the prominence has vanished, the intensity strength drops by an order of magnitude {for 171 \AA\ and almost two orders of magnitude for 304 \AA}. This trend is in very good agreement with the intensity evolution of \citet{wang2016}. Because the trend of our calculated intensity values agree so well with the observed intensities of \citet{wang2016}, we have the observational verification of our results.

Now that we have established that we indeed simulate an evaporating prominence, we can extend our discussion to the cause of prominence evaporation. In the last panel of the same figure, we display all the fluxes from the energy equation, except for the gravitational potential energy rate since it is negative in the demarcated region of Fig. \ref{fig:track_rectangle}. Note that we show the absolute value of the radiative cooling rate (identified by the thick black line). Only those fluxes that are positive and greater than the absolute value of the radiative cooling rate can heat up the prominence until evaporation. We immediately notice that Ohmic heating (blue line) and our static background heating (purple line) are much smaller in magnitude than the radiative cooling rate and hence cannot evaporate a prominence. Although the Poynting flux (beige line) occasionally reaches values similar to that of the radiative cooling rate, it is almost always weaker than radiative cooling\footnote{Since we disabled footpoint driving in the bottom boundary, there is no significant transport of Poynting flux from the bottom boundary to the simulation domain. However, when footpoint driving is active, it may influence the importance of Poynting flux in prominence evaporation.}. Only thermal conduction (red line) consistently dominates over radiative cooling throughout the entire evaporation phase at a value of $\sim 10^{-2} \ \mathrm{erg} \, \mathrm{cm}^{-3} \, \mathrm{s}^{-1}$. Additionally, compressional heating (green line) is also comparable to radiative cooling and sometimes exceeds it in magnitude. After the prominence has evaporated, thermal conduction decreases in magnitude whereas compressional heating maintains a significantly high value of $\sim 10^{-2} \ \mathrm{erg} \, \mathrm{cm}^{-3} \, \mathrm{s}^{-1}$. This value is almost in perfect agreement with \citet{lee2017}, who estimated the heating rate of an erupting prominence at $0.05 - 0.08 \ \mathrm{erg} \, \mathrm{cm}^{-3} \, \mathrm{s}^{-1}$. Though, it is important to note that in their article, the erupting prominence did not evaporate. Nevertheless, we conclude that thermal conduction and compressional heating are the important drivers in evaporating a solar prominence. They can further be relevant to erupting filaments that undergo heating without evaporation as well, which could explain the heated erupting filaments as observed by \citet{landi2010} and \citet{lee2017}. Moreover, \citet{vale} modelled an observed CME through data-driven simulations and showed that compression leads to the formation of the flux rope as well as its heating.  
We note that this result on prominence evaporation due to thermal conduction and compressional {heating is, to our knowledge, the first simulation-based study that explicitly demonstrates prominence }evaporation during the CME eruption.

\begin{figure}[thp]
    \includegraphics[width=0.48\textwidth]{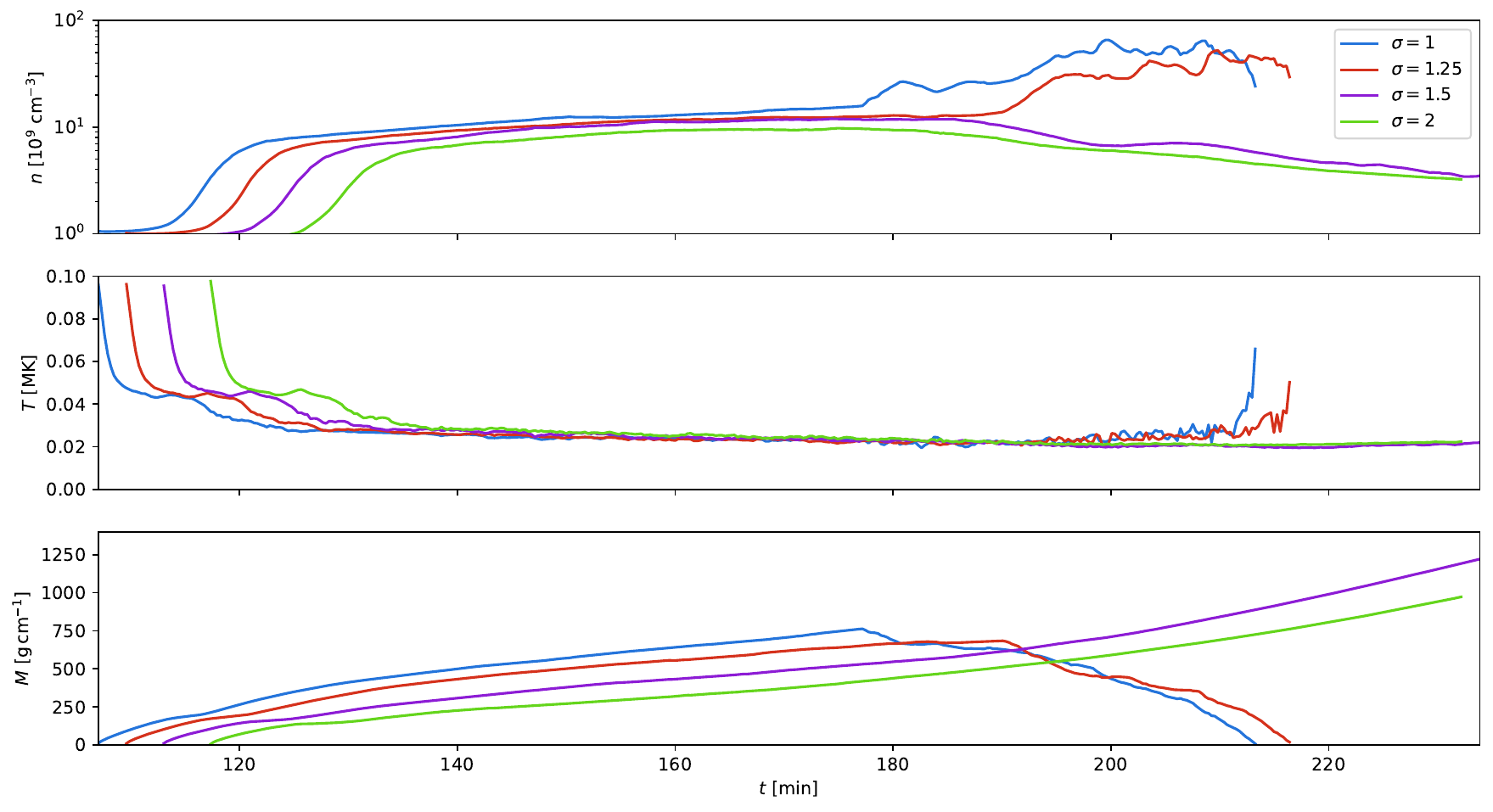}
    \caption{Average density $n$ in cubic centimetres (top panel), average temperature $T$ in megakelvin (middle panel) and total mass of the prominence $M$ in grams per centimetre (bottom panel) for the erupting case $\sigma \in\{1, 1.25, 1.5, 2\}$. The prominence is defined as regions where $T \leqslant 0.1 \ \mathrm{MK}$.} \label{fig:average_prom}
\end{figure}

To provide a meaningful comparison of the prominence among the four erupting cases, we average over regions $T \leqslant 0.1 \ \mathrm{MK}$ for the density and temperature and calculate the total mass of the prominence. This comparison is provided by Fig. \ref{fig:average_prom}. There, it can be seen that higher energy states $\sigma$ postpone the formation of the solar prominence. This agrees with our earlier observation that {the temperature of the flux rope and its ambient environment increases with increasing $\sigma$. Therefore, since the temperature of the flux rope increases with increasing $\sigma$, radiative cooling requires more time to cool down from the hotter flux rope temperatures to the prominence temperature of $T_\mathrm{min} \approx 0.1 \ \mathrm{MK}$ as can be seen from all the panels.} The average density grows rapidly for all curves and subsequently grows very slowly. For the blue and red curves ($\sigma\in\{1,1.25\}$), the average density increases substantially after  $t \gtrsim 176 \ \mathrm{min}$, respectively. That the average density increases is in agreement with our earlier work \citep{donne2024} (see discussion related to their Fig. 3): there, we argued that most of the prominence mass, undergoing minor evaporation, is stored as cold condensations $T \leqslant 0.025 \ \mathrm{MK}$. In the current case, since these denser condensations lie deeply within the prominence and are shielded by less dense condensations $0.025  \ \mathrm{MK} < T \leqslant 0.1 \ \mathrm{MK}$, the less dense condensations are evaporated first, resulting in that more dense condensations $T < 0.025  \ \mathrm{MK}$ are left behind and hence giving rise to the increase in the average density $n$ for both curves. This argument is further confirmed by the plot of the average temperature in the second panel of the same figure which shows that as the prominence undergoes evaporation for  $176 \ \mathrm{min} \leqslant t$ and $190 \ \mathrm{min} \leqslant t$ for $\sigma = 1$ and $\sigma=1.25$, respectively, the average temperature lies still around $T \sim 0.025 \ \mathrm{MK}$ and only when thermal conduction and compressional heating have reached the denser condensations, does the average temperature rise again sharply at $t\approx 219 \ \mathrm{min}$ for $\sigma = 1$, respectively. For the purple ($\sigma=1.5$) and green ($\sigma=2$) curves, the average density decreases after roughly the same time $185 \ \mathrm{min} < t$, which marks the instant when the flux rope erupts for both cases, see Fig. \ref{fig:height_evolution}. Despite the decreasing density, the mass steadily increases for both curves as can be seen from the bottom panel of the same figure. The total mass of the red and blue curves diminish to zero due to evaporation and disappearance of the prominence all together.

\subsection{Observational similarities}
Having established main novelties in our parametric study of filament-CME eruptions and having made quantitative agreements with observations, we will now correlate our findings with direct satellite observations and highlight morphological similarities. We made use of the \textit{AIA Filament Eruption Catalog}\footnote{\url{https://aia.cfa.harvard.edu/filament/index.html}} \citep{mccauley2015} to find early evolutions of filament eruptions as observed by the SDO/AIA instrument \citep{sdoaia}.

\begin{figure}[thp]
    \includegraphics[width=0.48\textwidth]{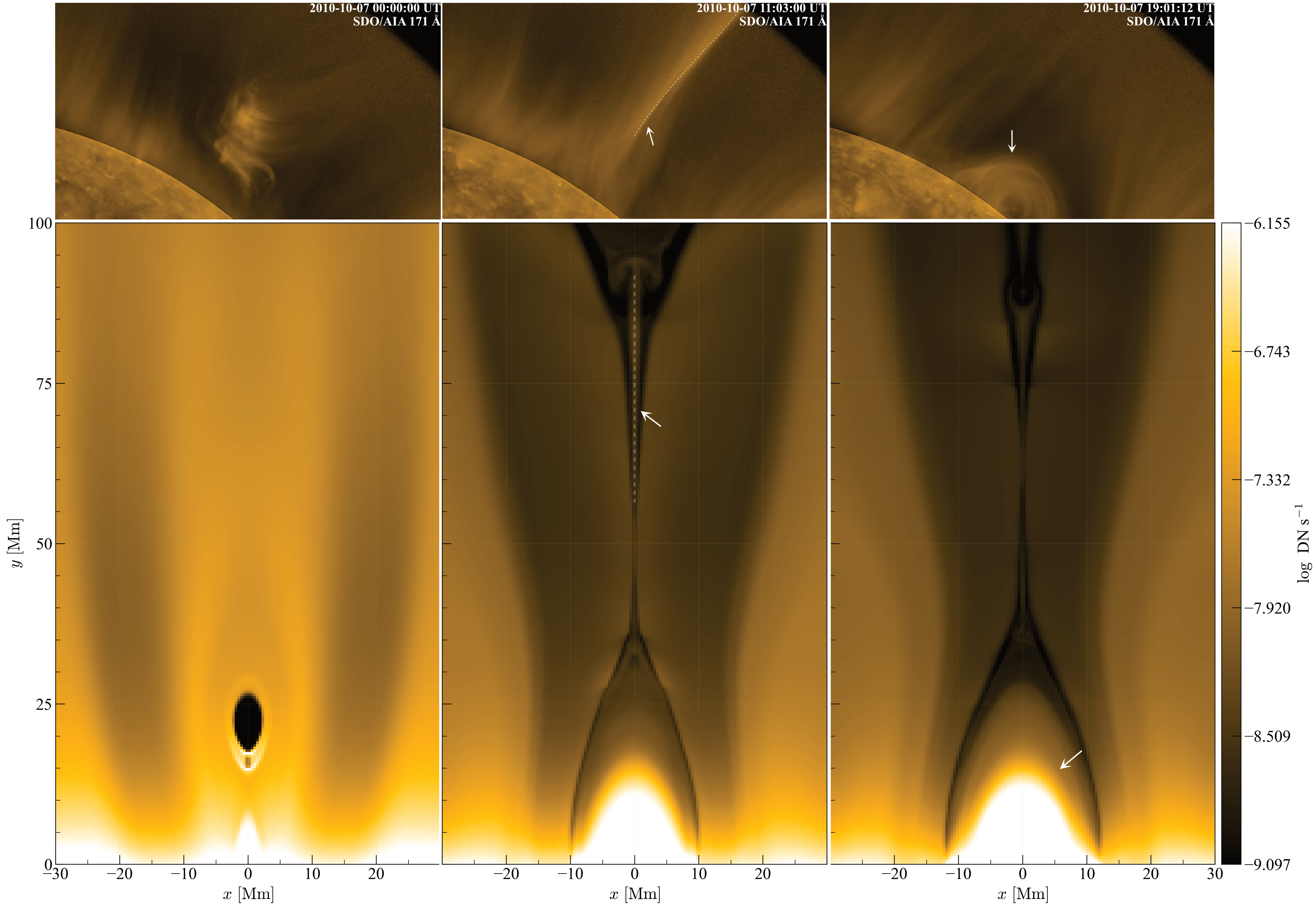}
    \caption{Top row shows the observed evolution of a filament eruption in the 171 \AA \ channel. In the top row, first panel shows the filament before the eruption. After it has erupted, a bright fan can be seen in the second panel (indicated by the arrow and dashed line) which disappears eventually and shows post-eruptive arcades as can be seen in the third panel, indicated by the white arrow. Bottom row are our simulated results for $\sigma=1$ in the synthetic 171 \AA \ channel and at instrument resolution, highlighting the same structures with white arrows and dashed lines as well. Observational images are retrieved from the AIA Filament Eruption Catalogue.} \label{fig:heatedfan}
\end{figure}

Fig. \ref{fig:heatedfan} shows an observed filament eruption on 07/10/2010 located in the NW limb and seen through the 171 \AA \ channel, manifesting a bright fan and after some time disappearing and giving rise to a conspicuous cusp structure and post-eruptive arcades, which almost always forms after CME events \citep{tripathi2004}. Our synthetic results in the same channel for $\sigma=1$ complement this evolution where we obtain the so-called fan structure which disappears eventually as well. Our results show that this fan structure corresponds to the current sheet where the density is $n\sim 3\cdot 10^8 \ \mathrm{cm}^{-3}$ and temperatures are $T\sim 2 \ \mathrm{MK}$. When the fan disappears, the density stays roughly the same whereas the temperature reaches values up to $T \sim 4 \ \mathrm{MK}$, which results in an order of magnitude decrease in the intensity compared to temperatures at $T\sim 2 \ \mathrm{MK}$ due to the sensitivity of the temperature response function of 171 \AA. After the observed prominence has erupted and the fan has disappeared, a dark cusp-shaped region forms which then gives way to post-eruptive arcades. In our case, the cusp-shaped region is retrieved and forms simultaneously with the post-eruptive arcades that have temperatures around $T \sim 1.3 \ \mathrm{MK}$.

\begin{figure}[thp]
    \includegraphics[width=0.23\textwidth]{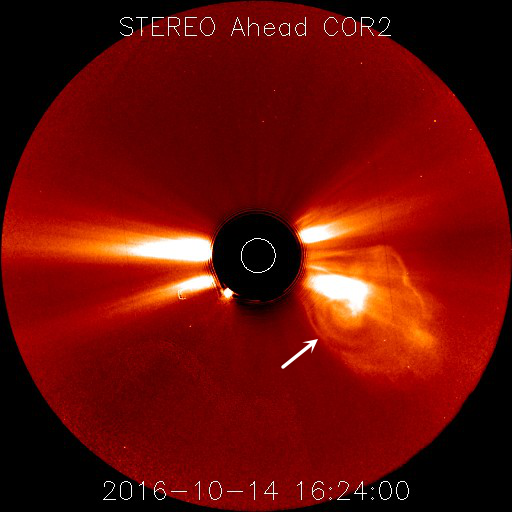} 
    \includegraphics[width=0.23\textwidth]{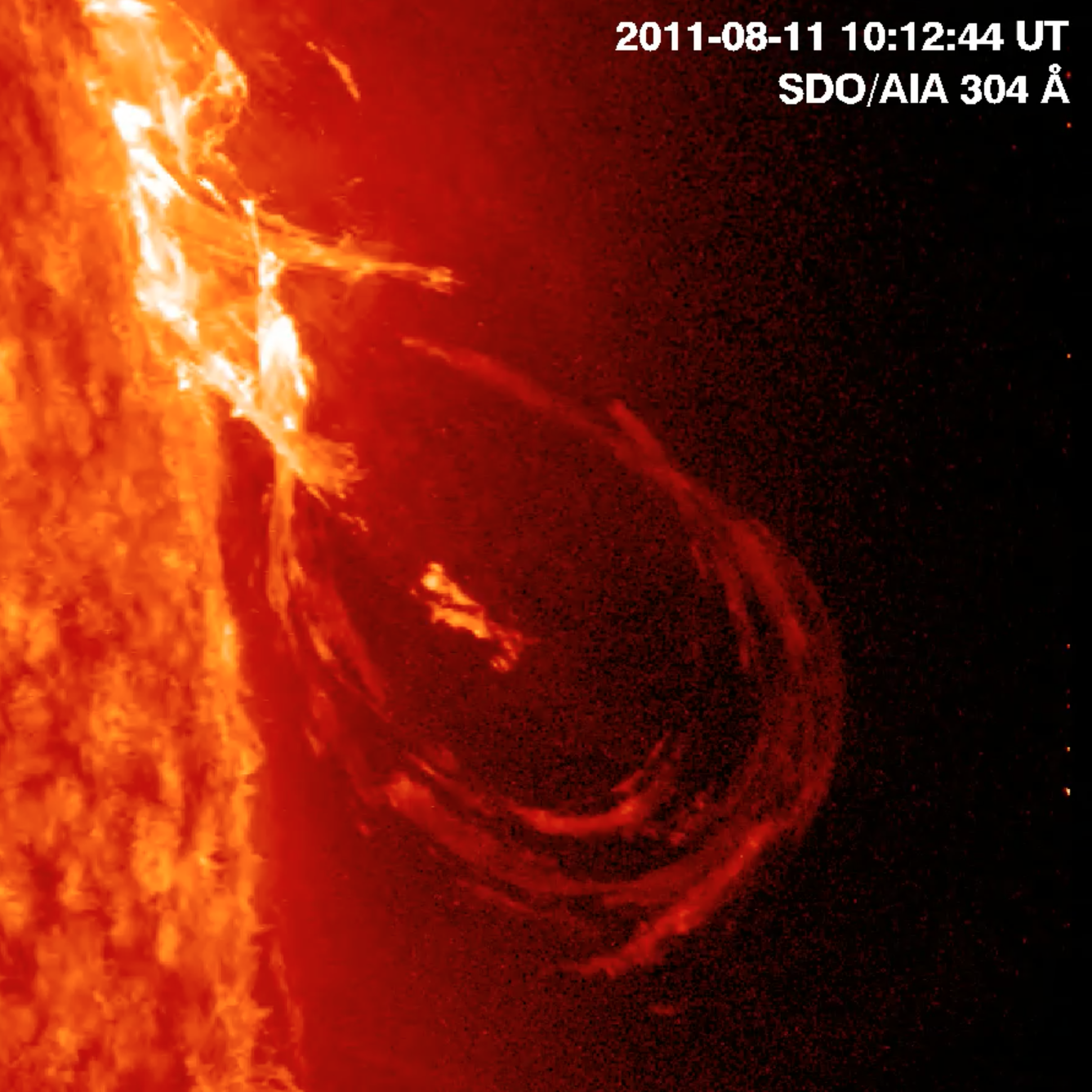} \\
    \includegraphics[width=0.23\textwidth]{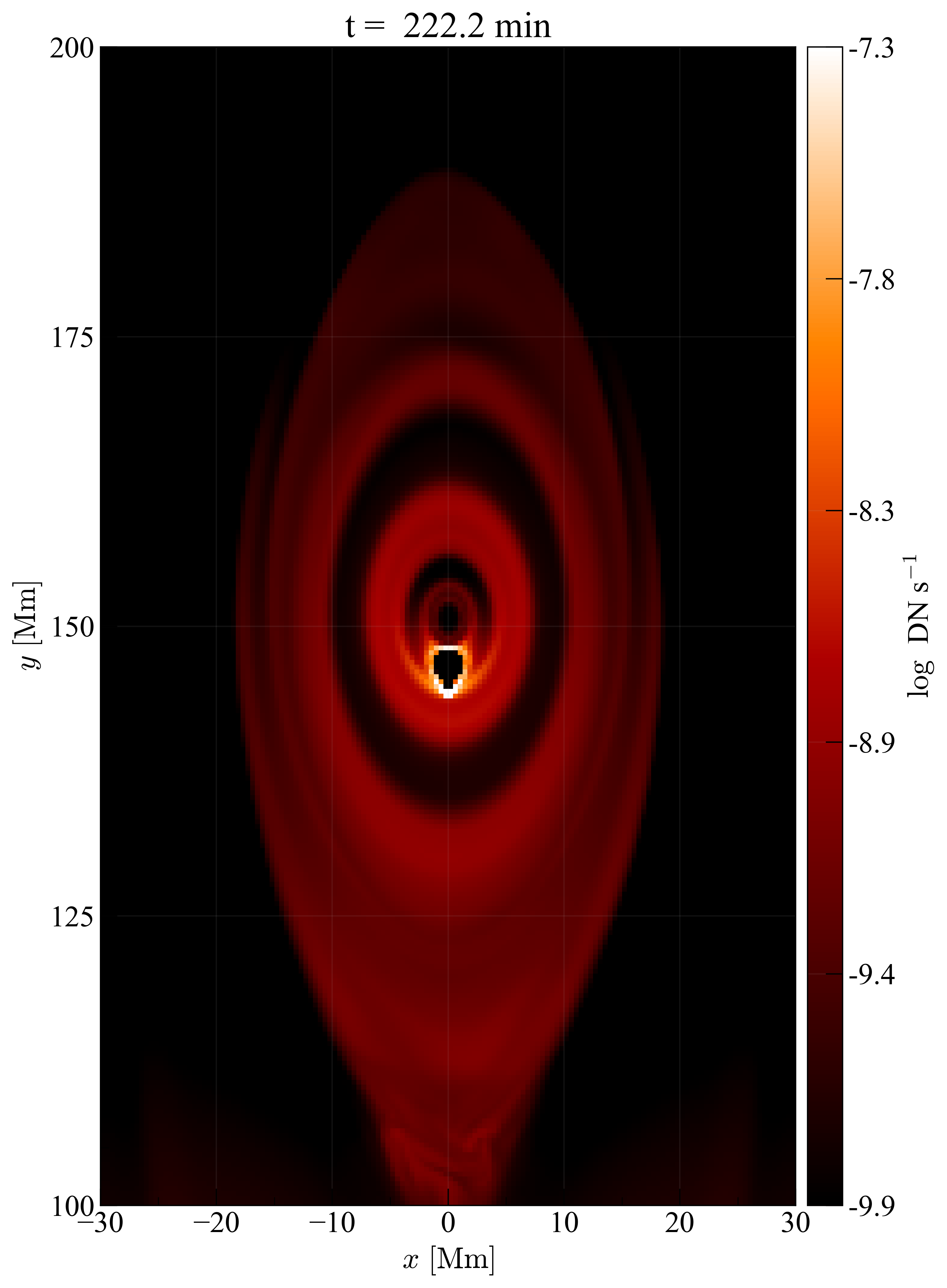}
    \includegraphics[width=0.23\textwidth]{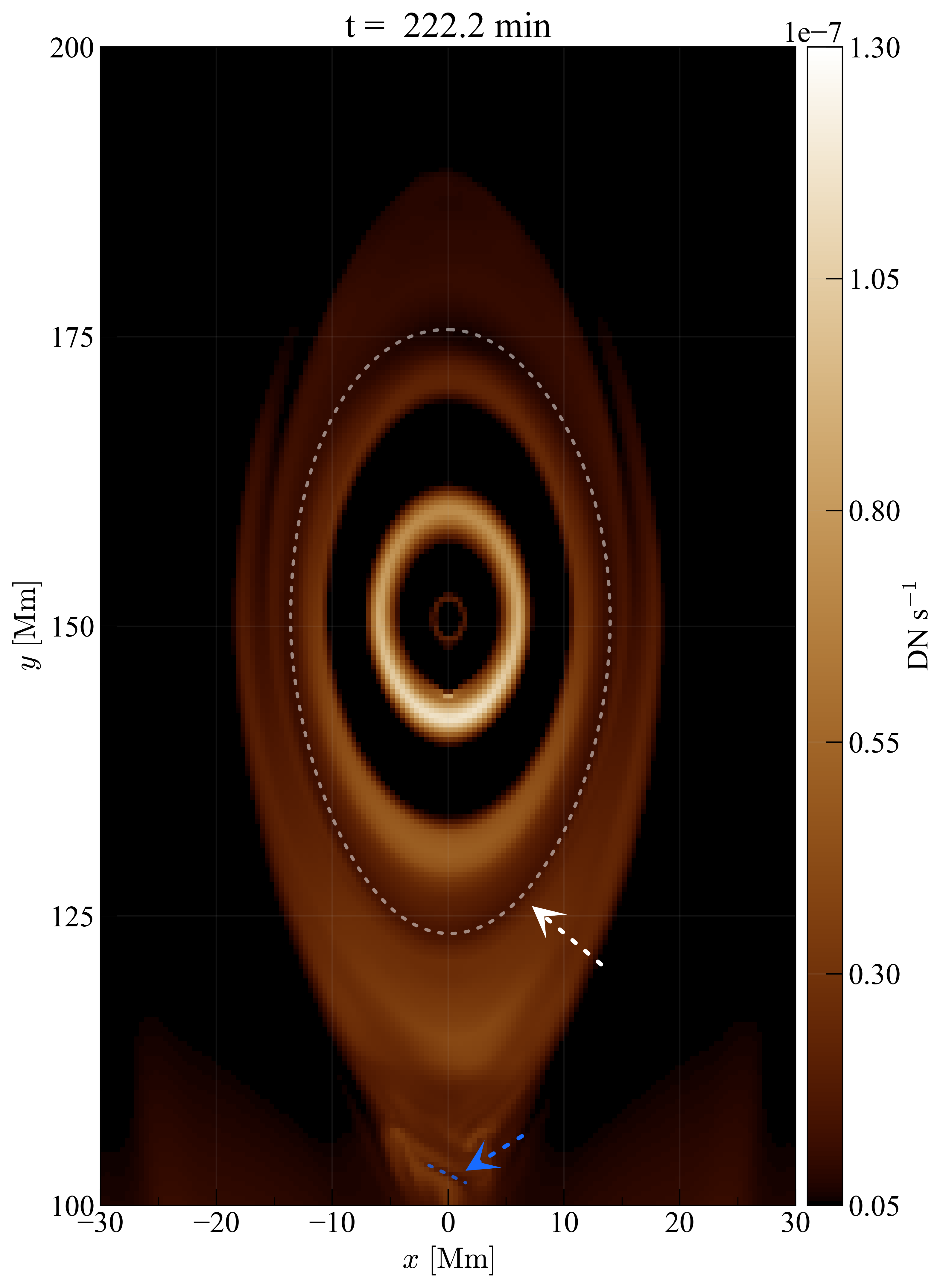}
    \caption{Top row display separate observed events through different satellites. Top left is a whitelight image from STEREO-A Cor2 of an erupting filament, displaying a clear circular pattern. Top right is a different event as observed through SDO/AIA in the 304 \AA \ channel with also a clear circular pattern.  Bottom row shows our synthetic images in the 304 \AA \ channel (bottom left panel) and 193 \AA \ channel (bottom right panel) for the $\sigma=2$ case at instrument resolution. {The blue and white arrows point to dark structures (that are in addition marked by the same-coloured, dashed lines) that are also found in Fig. \ref{fig:circular_pattern_analysis}.} SDO/AIA images are retrieved from the AIA Filament Eruption Catalogue.} \label{fig:circular_pattern}
\end{figure}

In Fig. \ref{fig:circular_pattern} we exhibit two different observations of erupting prominences with a strong circular pattern viewed from different satellites, and our complementing synthetic images. The top left panel is a whitelight image taken from STEREO-A COR2, showing the CME with a semi-bright core surrounded by a small dark gap, then a bright ring, then a gap and again a bright tiny ring. Since intensity in white light images depend on the local density of the electrons \citep{howard2009}, these bright rings can indicate regions of higher local electron density and gap regions with lower density. In the top right panel, we show a separate event as imaged by SDO/AIA in the 304 \AA \ channel where this circular pattern is more outspoken. This indicates that fine structure such as the circular pattern can be detected in the low corona. In the bottom left- and right panel we have made synthesized images in the 304 \ \AA \ and 193 \AA \ channel for the $\sigma=2$ case. They indeed mirror the circular pattern as observed.

\begin{figure}[thp]
\includegraphics[width=0.48\textwidth]{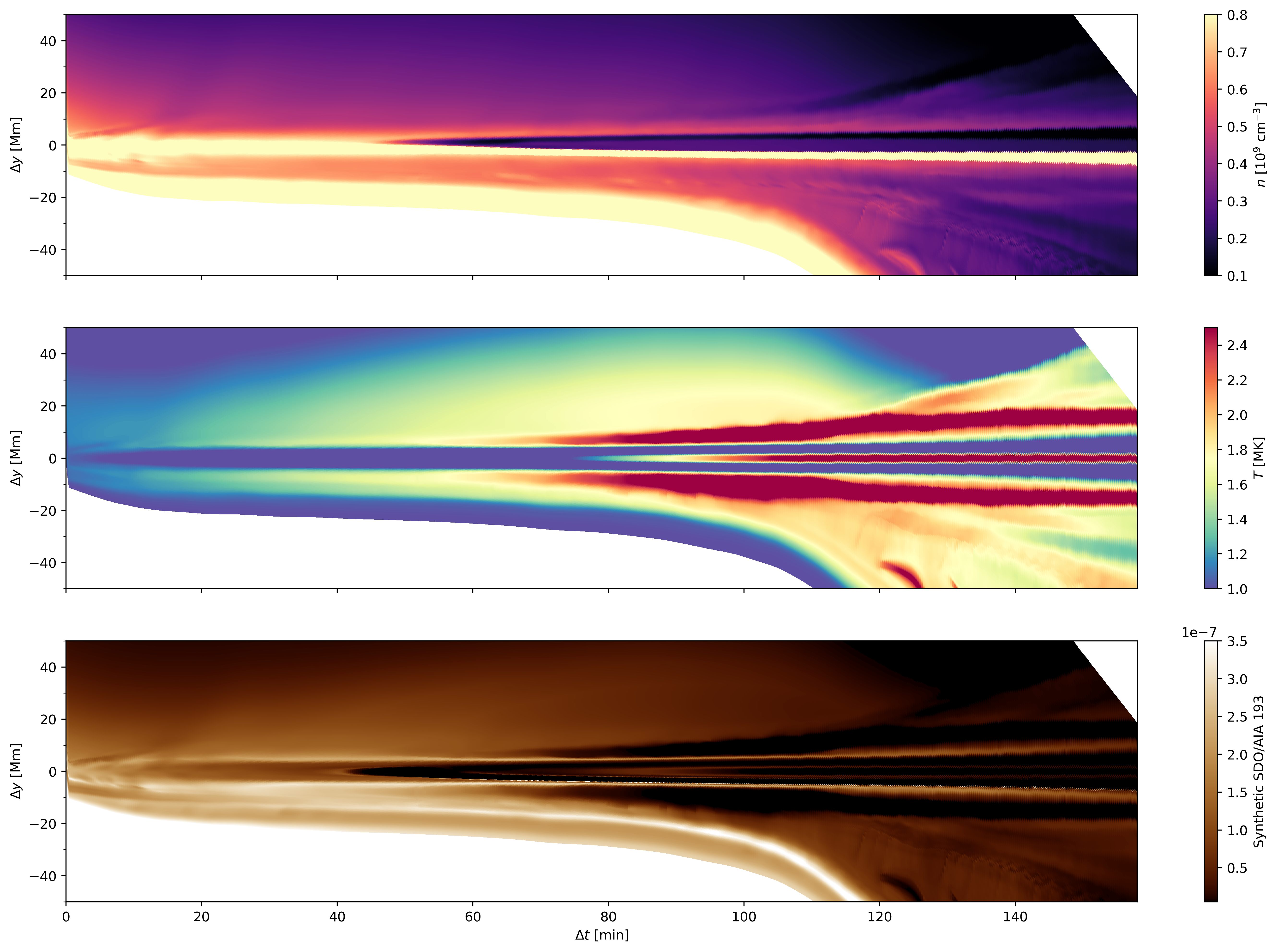}
\caption{Evolution of the density in $10^9 \ \mathrm{cm}^{-3}$ (top panel), temperature in MK (middle panel) and intensity in the synthetic 193 \AA \ channel (bottom panel) along the central vertical axis $x=0$ in the reference frame of the flux rope centre. $\Delta \, y$ indicates the height with respect to the flux rope centre and $\Delta \, t$ the time with respect to the moment when the flux rope centre is tracked.}\label{fig:circular_pattern_analysis}
\end{figure}

We have done an analysis regarding the origin of the circular pattern and the results can be seen in Fig.~\ref{fig:circular_pattern_analysis}. 
The three panels show the evolution along the central vertical axis $x=0$ for the density, temperature and intensity of the synthesized 193 \AA\ channel in the reference frame of the flux rope centre. Hence, $\Delta y = 0$ indicates the centre of the flux rope. 
Several darker and brighter regions can be identified in the synthetic emission within the flux rope. 
By comparing the density and temperature evolution with the synthetic intensity, we find that the darker rings or gaps can be caused by extreme temperatures, either $T<0.1\,\mathrm{MK}$ where the prominence is located or $T>2.5\,\mathrm{MK}$ where the density is roughly $n\sim0.25\times10^9\,\mathrm{cm}^{-3}$. 

In contrast, the bright regions have roughly double the density, $n\sim0.5\times10^9\,\mathrm{cm}^{-3}$ compared to the average density within the flux rope, and temperatures around $T\sim1.5\,\mathrm{MK}$. The bright circular pattern is therefore a combination of enhanced density and a suitable temperature, since the temperature response function of the 193 \AA\ channel peaks at $10^{6.2}\,\mathrm{K}\approx1.6\,\mathrm{MK}$. The darker rings can thus be either hot, cold, sparse, or dense.

Interestingly, plasmoids can also give rise to darker rings in the synthetic image. At certain times, plasmoids interact with the flux rope, which subsequently leads to a local darkening in the synthetic intensity after a short delay, creating a dark gap. While the density in this gap can be relatively high ($n\sim0.6\times10^9\,\mathrm{cm}^{-3}$), the temperature exceeds $2\,\mathrm{MK}$, where the temperature response function is significantly reduced. 
We also note the presence of additional dark structures that recede from the flux rope centre. These features correspond to the interaction interface between the erupting flux rope and the plasmoids generated in the current sheet.

\begin{figure}[thp]
    \begin{center}
    \includegraphics[width=0.235\textwidth]{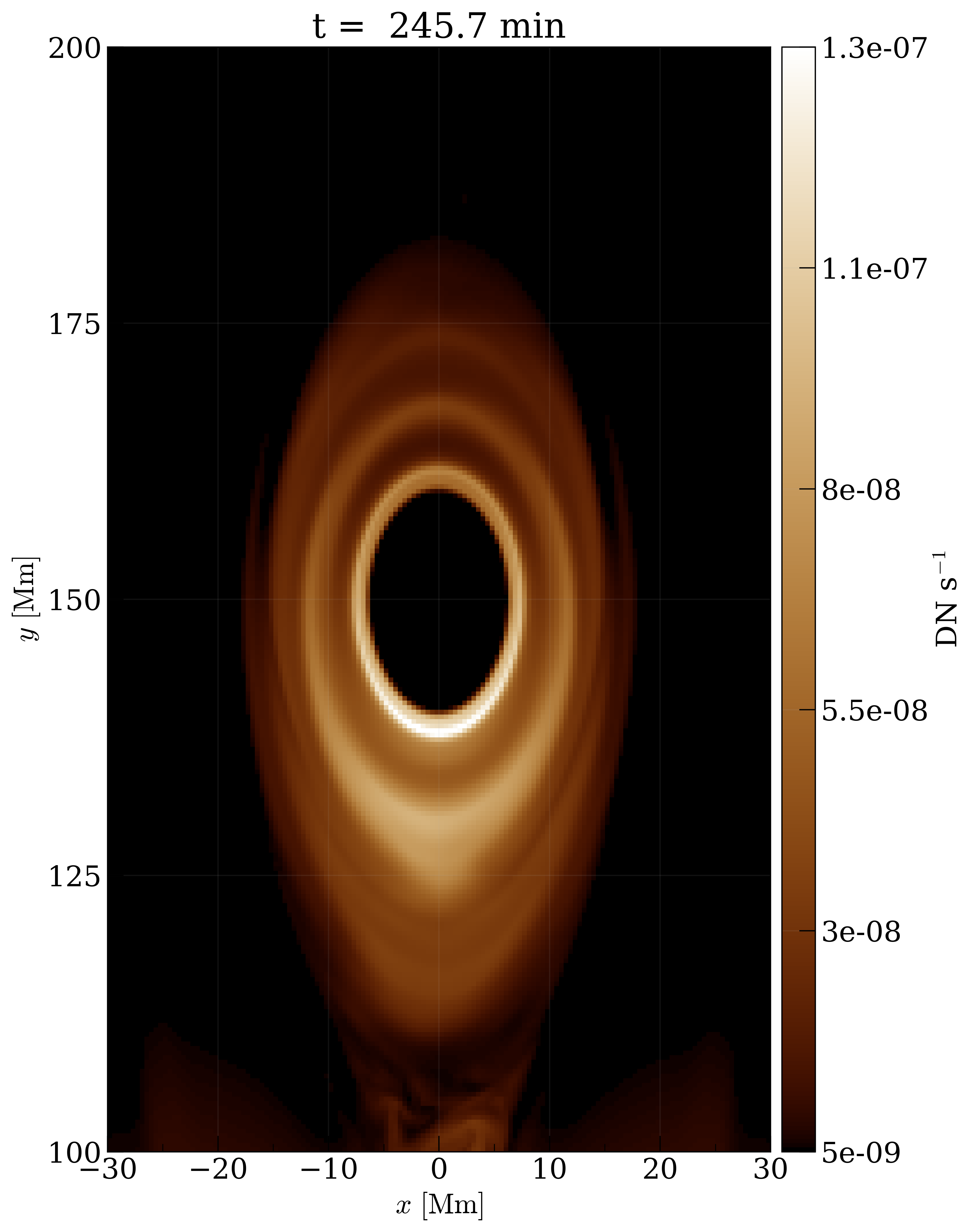}
    \includegraphics[width=0.235\textwidth]{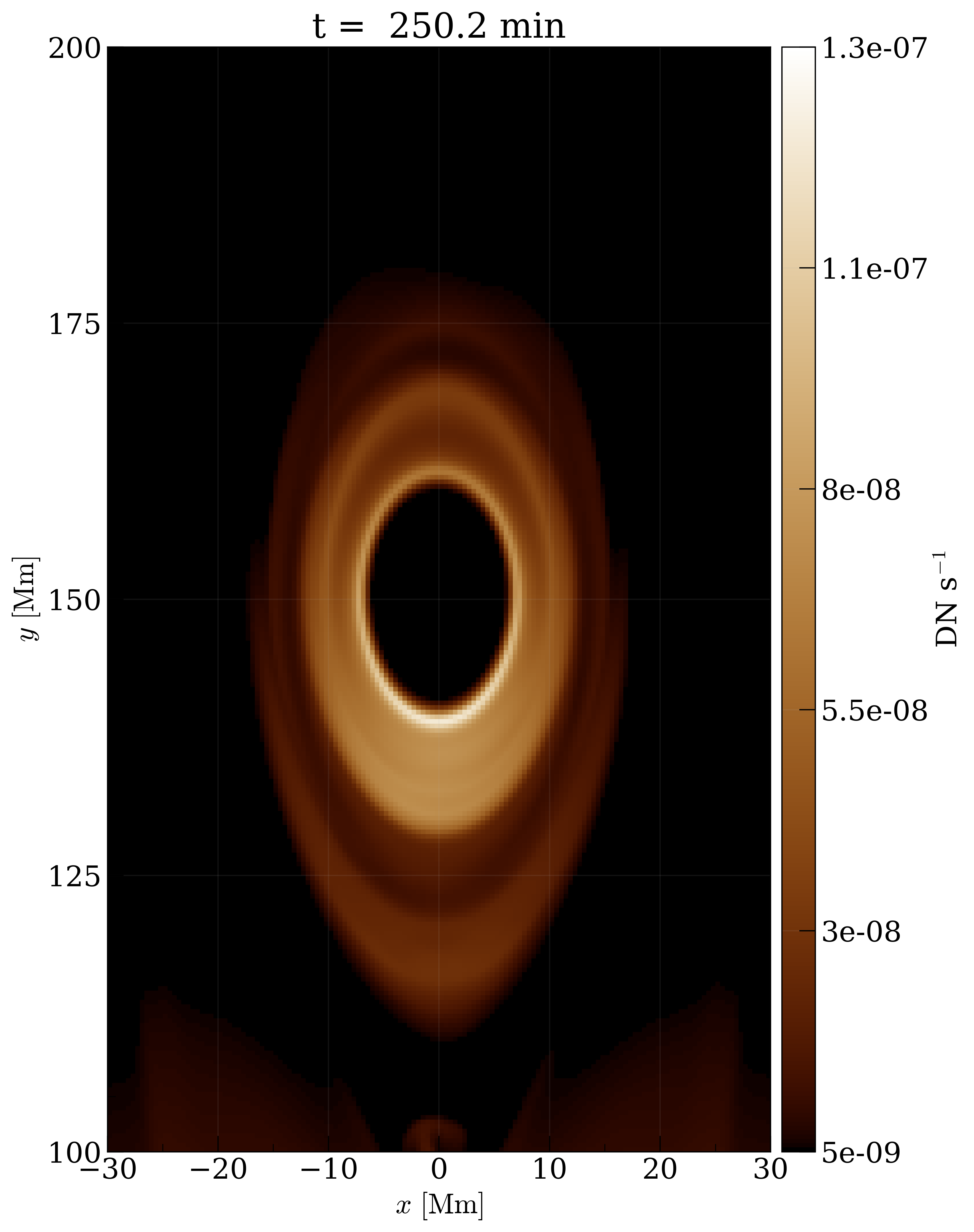}
    \includegraphics[width=0.235\textwidth]{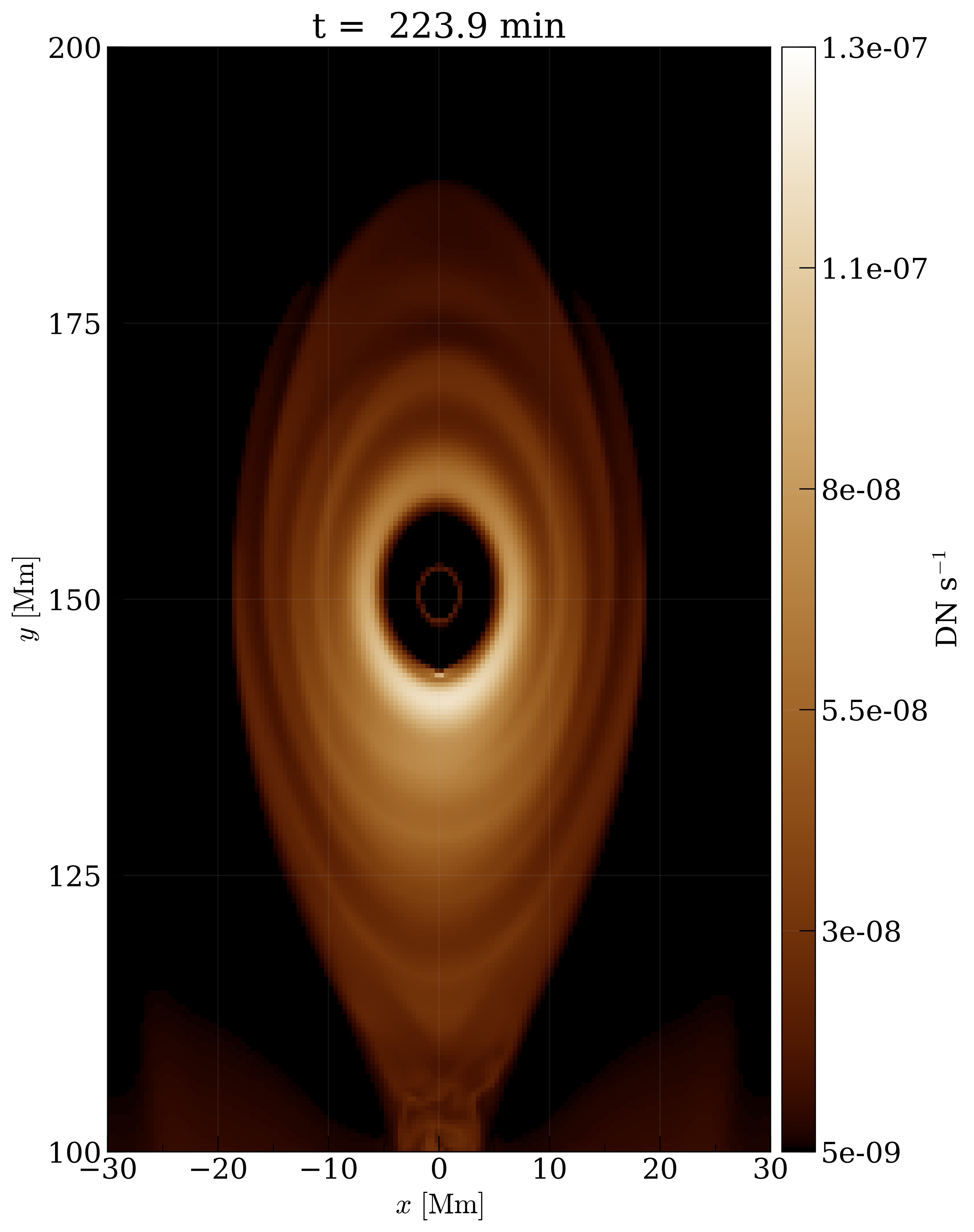}
    \includegraphics[width=0.235\textwidth]{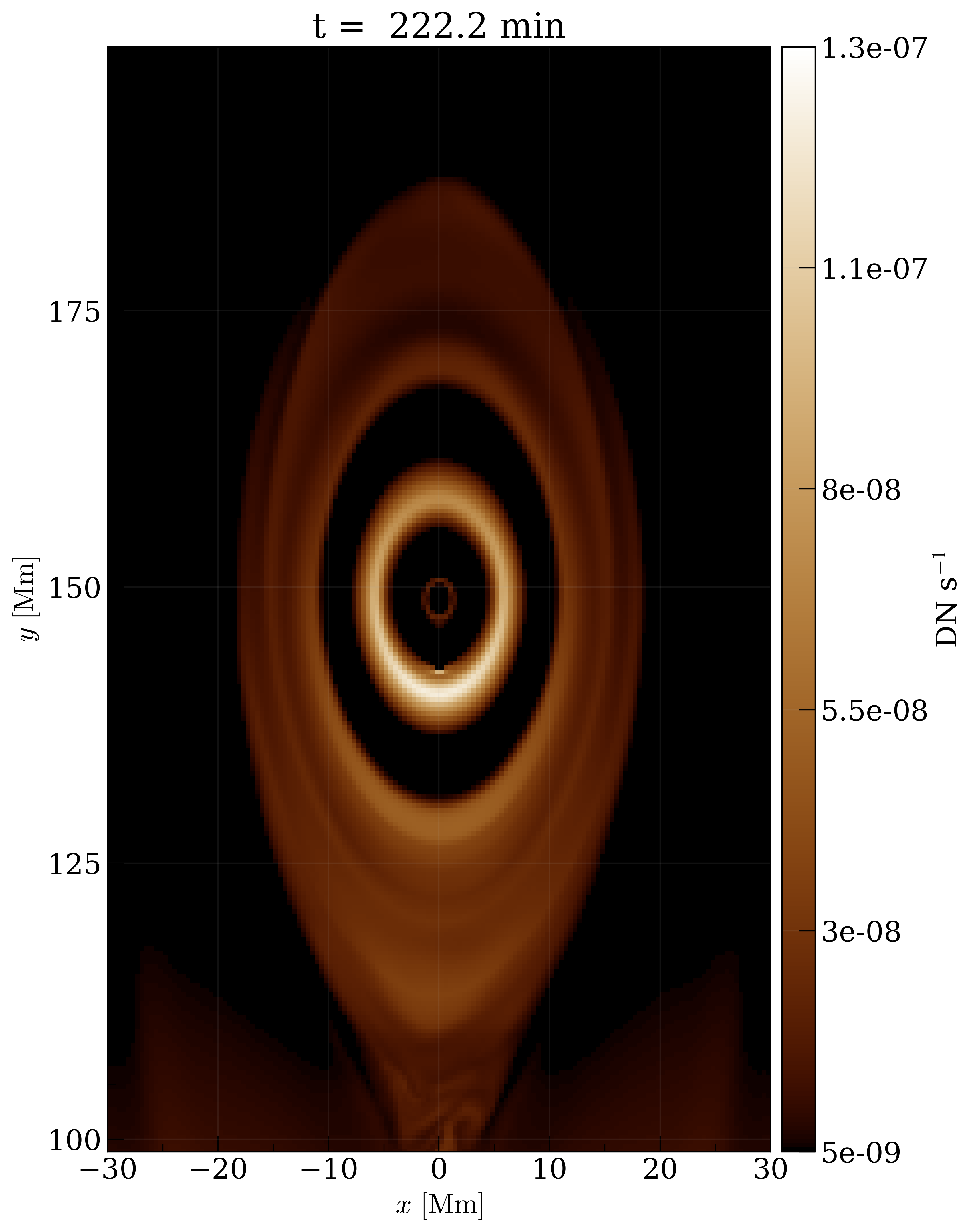}
    \caption{Morphology of the erupting flux ropes in the synthetic 193 \AA \ channel for $\sigma=1$ (top left), $\sigma=1.25$ (top right), $\sigma=1.5$ (bottom left) and $\sigma=2$ (bottom right).} \label{fig:morphology_cmes}
    \end{center}
\end{figure}

While plasmoids may contribute to darker gaps in the synthetic images, it is not a sufficient condition. Fig. \ref{fig:morphology_cmes} shows the different morphologies of the erupting flux ropes for the different erupting cases $\sigma$. All erupting cases $\sigma \in \{1, 1.25, 1.5, 2\}$ form plasmoids but only $\sigma=2$ shows a very strong annular pattern in the synthetic image. The case of $\sigma=1.25$ has a small dark ring but not as conspicuous as $\sigma=2$. This indicates that the morphology of erupting flux ropes are very intricate to the energy state of the flux rope, considering that the largest relative difference in maximum magnetic energy of each case $\sigma$ accounts for $2.1\%$. 

\section{Summary and conclusion}
We have examined the response of the flux rope to only slightly varying its energy state, i.e. increased shearing during flux rope formation. The largest relative difference between the maximally attained magnetic energy amongst the different shearing cases $\sigma$ amounts to only 2.1\% and, despite this low value, our erupting prominences have shown a rich variety in phenomena. 

For the case $\sigma=0$ the flux rope does not show any sign of eruption whereas for $\sigma=0.75$ the flux rope centre slightly increases in altitude and because of its continuously decreasing magnetic energy, we speculate that it will eventually erupt as well if our simulation time were to be extended. The cases $\sigma \in\{1, 1.25, 1.5, 2\}$ do erupt. From the quantifications in Table \ref{table:energy_conversion}, higher energy states of the flux rope erupt faster. Though, a one-to-one correspondence between higher energy states $\sigma$ and higher conversion rates to kinetic and thermal energy does not exist. Furthermore, we obtain the slow-rise phase of the CME for the erupting cases. \citet{liu2025} owed this slow rise phase to the expansion of the overlying arcade field, but our results show that this is not a necessary condition as our overlying arcades do not expand noticeably: magnetic reconnection itself can reproduce the slow-rise phase, in agreement with \citet{fan2017,xing2024}. 
We do acknowledge that our 2.5D setup only captures what happens centrally in a full 3D erupting flux rope. 
{For example, 2.5D simulations cannot capture prominence draining along field lines, which in fully 3D configurations can contribute to or even drive the slow-rise phase, as suggested by \citet{fan2020} and demonstrated by \citet{xing2025}. 
In this context, }We proved that the condition for prominence eruption in our case lies with magnetic reconnection: the inflow Alfvén Mach number increased by an order of magnitude with respect to the slow rise phase.

When we compared the thermodynamics of the solar prominences, we have found that for $\sigma\in\{1,1.25\}$ the prominence evaporates. We have proven that this evaporation is due to thermal conduction and compressional heating. When we quantified the average synthetic intensities in the 304 \AA \ and 171 \AA \ channels of the prominence evaporation, we find excellent agreement with an observed prominence evaporation \citep{wang2016}. This unique finding, which has never been demonstrated before in the literature, showed that thermal conduction and compressional heating are very relevant processes in not only evaporating solar prominences, but also heating the flux rope interior. For the cases $\sigma=\{1.5, 2\}$ the prominence continues to exist throughout eruption. These results verify that flux rope interiors are unique and susceptible to strongly varying thermodynamics, as motivated by \citet{low2001}. It would be interesting to extend our simulations of $\sigma\in \{1, 1.25 \}$ to 3D and examine whether the prominence evaporation would still persist. If the latter were to be the case and it would indeed endure these dominant fluxes, our idealised setup could be a strong model candidate for the event observed by \citet{lee2017}, considering that our heating rates of our erupting prominences already agree very well in 2.5D with their observed filament erupting undergoing heating. This would then further support our argument for the clear thermodynamic role of thermal conduction and compressional heating in prominence eruptions.

We finalised our work by drawing parallels with observations. We obtained a fan structure after the prominences have erupted. In addition, for the first time as well, fine structure such as a circular pattern has been retrieved in our synthetic (EUV) data for $\sigma=2$ which has been observed in white light images \citep{dere1999, vourlidas2013}. The bright intensity rings are due to a relatively high density compared to other regions in the flux rope and temperatures that match with the peak values of SDO/AIA's temperature response functions. About the darker gaps/rings in this circular pattern: they can indicate hot, cold, dense and/or sparse regions. For the first time as well, we showed that plasmoids can contribute to the fine structure of CMEs by causing a dark ring in this circular pattern. This is not a sufficient condition, however, as our other $\sigma$-cases form plasmoids as well but do not always possess a (noticeable) circular pattern in their synthetic images. From these results we infer that fine structures in the white light in the heliosphere can already pre-exist in the low corona. More research is needed to validate this hypothesis by extending the physical domain up to heliospheric length scales and examine whether these fine structures will still persist throughout the CME evolution.

For a follow-up study, we aim to realise a 3D model of our prominence eruptions. By further changing to spherical coordinates and extending the top boundary to 10 solar radii, it will allow for synthetic white light images and facilitate comparisons with white light observations. Whereas in 2.5D the present study is constrained to plane-of-the-sky view, 3D simulations will, in addition, elucidate on CME morphology through different angles and investigate any projections effects. 

%--------------------------------------------------------------------

\begin{acknowledgements}
      {The authors want to thank the anonymous referee for the valuable feedback which has improved the quality of this article.} DD and HC acknowledge that the research was sponsored by the DynaSun project and has thus received funding under the Horizon Europe programme of the European Union under grant agreement (no. 101131534). Views and opinions expressed are however those of the author(s) only and do not necessarily reflect those of the European Union and therefore the European Union cannot be held responsible for them. The computational resources and services used in this work were provided by the VSC (Flemish Supercomputer Center), funded by the Research Foundation Flanders (FWO) and the Flemish Government, department EWI. 
       RK acknowledges funding from the KU Leuven C1 project C16/24/010 UnderRadioSun and the Research Foundation Flanders FWO project G0B9923N Helioskill. 
       YZ acknowledges the support from the Jiangsu Provincial Double First-Class Initiative (grant number 1480604106).
\end{acknowledgements}

\bibliographystyle{aa} % style aa.bst
\bibliography{bibliography} % your references Yourfile.bib

@article{sdoaia,
  title={The atmospheric imaging assembly (AIA) on the solar dynamics observatory (SDO)},
  author={Lemen, James R and Title, Alan M and Akin, David J and Boerner, Paul F and Chou, Catherine and Drake, Jerry F and Duncan, Dexter W and Edwards, Christopher G and Friedlaender, Frank M and Heyman, Gary F and others},
  journal={SoPh},
  volume={275},
  pages={17--40},
  year={2012},
  publisher={Springer}
}

@ARTICLE{amrvac3.0,
       author = {{Keppens}, R. and {Popescu Braileanu}, B. and {Zhou}, Y. and {Ruan}, W. and {Xia}, C. and {Guo}, Y. and {Claes}, N. and {Bacchini}, F.},
        title = "{MPI-AMRVAC 3.0: Updates to an open-source simulation framework}",
      journal = {\aap},
         year = 2023,
        month = may,
       volume = {673},
          eid = {A66},
        pages = {A66},
          doi = {10.1051/0004-6361/202245359},
archivePrefix = {arXiv},
       eprint = {2303.03026},
 primaryClass = {astro-ph.IM},
       adsurl = {https://ui.adsabs.harvard.edu/abs/2023A&A...673A..66K}
}

@ARTICLE{amari2000,
       author = {{Amari}, T. and {Luciani}, J.~F. and {Mikic}, Z. and {Linker}, J.},
        title = "{A Twisted Flux Rope Model for Coronal Mass Ejections and Two-Ribbon Flares}",
      journal = {\apjl},
         year = 2000,
        month = jan,
       volume = {529},
       number = {1},
        pages = {L49-L52},
          doi = {10.1086/312444},
       adsurl = {https://ui.adsabs.harvard.edu/abs/2000ApJ...529L..49A}
}

@ARTICLE{antiochos1999,
       author = {{Antiochos}, S.~K. and {DeVore}, C.~R. and {Klimchuk}, J.~A.},
        title = "{A Model for Solar Coronal Mass Ejections}",
      journal = {\apj},
         year = 1999,
        month = jan,
       volume = {510},
       number = {1},
        pages = {485-493},
          doi = {10.1086/306563},
archivePrefix = {arXiv},
       eprint = {astro-ph/9807220},
 primaryClass = {astro-ph},
       adsurl = {https://ui.adsabs.harvard.edu/abs/1999ApJ...510..485A}
}

@article{berger2010,
  title={Quiescent prominence dynamics observed with the Hinode solar optical telescope. I. Turbulent upflow plumes},
  author={Berger, Thomas E and Slater, Gregory and Hurlburt, Neal and Shine, Richard and Tarbell, Theodore and Lites, Bruce W and Okamoto, Takenori J and Ichimoto, Kiyoshi and Katsukawa, Yukio and Magara, Tetsuya and others},
  journal={\apj},
  volume={716},
  number={2},
  pages={1288},
  year={2010},
  publisher={IOP Publishing}
}

@article{brughmans2022,
  title={The influence of flux rope heating models on solar prominence formation},
  author={Brughmans, N and Jenkins, JM and Keppens, R},
  journal={\aap},
  volume={668},
  pages={A47},
  year={2022},
  publisher={EDP Sciences}
}

@ARTICLE{chen2011,
       author = {{Chen}, P.~F.},
        title = "{Coronal Mass Ejections: Models and Their Observational Basis}",
      journal = {Living Reviews in Solar Physics},
         year = 2011,
        month = dec,
       volume = {8},
       number = {1},
          eid = {1},
        pages = {1},
          doi = {10.12942/lrsp-2011-1},
       adsurl = {https://ui.adsabs.harvard.edu/abs/2011LRSP....8....1C}
}

@INPROCEEDINGS{chiu1976,
       author = {{Chiu}, Y.~T. and {Hilton}, H.~H.},
        title = "{Exact Green's Function Method of Solar Force-Free Magnetic Field Computations with Constant {\ensuremath{\alpha}}: Theory and Basic Test Cases}",
    booktitle = {Bulletin of the American Astronomical Society},
         year = 1976,
       volume = {8},
        month = mar,
        pages = {370},
       adsurl = {https://ui.adsabs.harvard.edu/abs/1976BAAS....8..370C}
}

@ARTICLE{delmoro2007,
       author = {{Del Moro}, D. and {Giordano}, S. and {Berrilli}, F.},
        title = "{3D photospheric velocity field of a supergranular cell}",
      journal = {\aap},
         year = 2007,
        month = sep,
       volume = {472},
       number = {2},
        pages = {599-605},
          doi = {10.1051/0004-6361:20077595},
archivePrefix = {arXiv},
       eprint = {0704.0578},
 primaryClass = {astro-ph},
       adsurl = {https://ui.adsabs.harvard.edu/abs/2007A&A...472..599D}
}

@ARTICLE{dere1999,
       author = {{Dere}, K.~P. and {Brueckner}, G.~E. and {Howard}, R.~A. and {Michels}, D.~J. and {Delaboudiniere}, J.~P.},
        title = "{LASCO and EIT Observations of Helical Structure in Coronal Mass Ejections}",
      journal = {\apj},
         year = 1999,
        month = may,
       volume = {516},
       number = {1},
        pages = {465-474},
          doi = {10.1086/307101},
       adsurl = {https://ui.adsabs.harvard.edu/abs/1999ApJ...516..465D}
}

@ARTICLE{donne2024,
       author = {{Donn{\'e}}, D. and {Keppens}, R.},
        title = "{Mass Cycle and Dynamics of a Virtual Quiescent Prominence}",
      journal = {\apj},
         year = 2024,
        month = aug,
       volume = {971},
       number = {1},
          eid = {90},
        pages = {90},
          doi = {10.3847/1538-4357/ad50a3},
archivePrefix = {arXiv},
       eprint = {2405.20048},
 primaryClass = {astro-ph.SR},
       adsurl = {https://ui.adsabs.harvard.edu/abs/2024ApJ...971...90D}
}

@ARTICLE{dorman2008,
       author = {{Dorman}, L.~I. and {Ptitsyna}, N.~G. and {Villoresi}, G. and {Kasinsky}, V.~V. and {Lyakhov}, N.~N. and {Tyasto}, M.~I.},
        title = "{Space storms as natural hazards}",
      journal = {Advances in Geosciences},
         year = 2008,
        month = apr,
       volume = {14},
        pages = {271-275},
          doi = {10.5194/adgeo-14-271-2008},
       adsurl = {https://ui.adsabs.harvard.edu/abs/2008AdG....14..271D}
}

@ARTICLE{fan2017,
       author = {{Fan}, Yuhong},
        title = "{MHD Simulations of the Eruption of Coronal Flux Ropes under Coronal Streamers}",
      journal = {\apj},
         year = 2017,
        month = jul,
       volume = {844},
       number = {1},
          eid = {26},
        pages = {26},
          doi = {10.3847/1538-4357/aa7a56},
archivePrefix = {arXiv},
       eprint = {1706.06076},
 primaryClass = {astro-ph.SR},
       adsurl = {https://ui.adsabs.harvard.edu/abs/2017ApJ...844...26F}
}

@ARTICLE{fan2018,
       author = {Fan, Yuhongho},
        title = "{MHD Simulation of Prominence Eruption}",
      journal = {\apj},
         year = 2018,
        month = jul,
       volume = {862},
       number = {1},
          eid = {54},
        pages = {54},
          doi = {10.3847/1538-4357/aaccee},
archivePrefix = {arXiv},
       eprint = {1806.06305},
 primaryClass = {astro-ph.SR},
       adsurl = {https://ui.adsabs.harvard.edu/abs/2018ApJ...862...54F}
}

@ARTICLE{fan2020,
       author = {{Fan}, Yuhong},
        title = "{Simulations of Prominence Eruption Preceded by Large-amplitude Longitudinal Oscillations and Draining}",
      journal = {\apj},
         year = 2020,
        month = jul,
       volume = {898},
       number = {1},
          eid = {34},
        pages = {34},
          doi = {10.3847/1538-4357/ab9d7f},
archivePrefix = {arXiv},
       eprint = {2006.11619},
 primaryClass = {astro-ph.SR},
       adsurl = {https://ui.adsabs.harvard.edu/abs/2020ApJ...898...34F}
}

@article{field1965,
       author = {{Field}, George B.},
        title = "{Thermal Instability.}",
      journal = {\apj},
         year = 1965,
        month = aug,
       volume = {142},
        pages = {531},
          doi = {10.1086/148317},
       adsurl = {https://ui.adsabs.harvard.edu/abs/1965ApJ...142..531F}
}

@ARTICLE{fillipov2002,
       author = {{Filippov}, B. and {Koutchmy}, S.},
        title = "{About the prominence heating mechanisms during its eruptive phase}",
      journal = {\solphys},
         year = 2002,
        month = aug,
       volume = {208},
       number = {2},
        pages = {283-295},
          doi = {10.1023/A:1020532607451},
       adsurl = {https://ui.adsabs.harvard.edu/abs/2002SoPh..208..283F}
}

@ARTICLE{forbes2000,
       author = {{Forbes}, T.~G.},
        title = "{A review on the genesis of coronal mass ejections}",
      journal = {\jgr},
         year = 2000,
        month = oct,
       volume = {105},
       number = {A10},
        pages = {23153-23166},
          doi = {10.1029/2000JA000005},
       adsurl = {https://ui.adsabs.harvard.edu/abs/2000JGR...10523153F}
}

@ARTICLE{georgoulis2019,
       author = {{Georgoulis}, Manolis K. and {Nindos}, Alexander and {Zhang}, Hongqi},
        title = "{The source and engine of coronal mass ejections}",
      journal = {Philosophical Transactions of the Royal Society of London Series A},
         year = 2019,
        month = jul,
       volume = {377},
       number = {2148},
        pages = {20180094},
          doi = {10.1098/rsta.2018.0094},
       adsurl = {https://ui.adsabs.harvard.edu/abs/2019RSPTA.37780094G}
}

@ARTICLE{gibson1998,
       author = {{Gibson}, S.~E. and {Low}, B.~C.},
        title = "{A Time-Dependent Three-Dimensional Magnetohydrodynamic Model of the Coronal Mass Ejection}",
      journal = {\apj},
         year = 1998,
        month = jan,
       volume = {493},
       number = {1},
        pages = {460-473},
          doi = {10.1086/305107},
       adsurl = {https://ui.adsabs.harvard.edu/abs/1998ApJ...493..460G}
}

@ARTICLE{glesener2013,
       author = {{Glesener}, Lindsay and {Krucker}, S{\"a}m and {Bain}, Hazel M. and {Lin}, Robert P.},
        title = "{Observation of Heating by Flare-accelerated Electrons in a Solar Coronal Mass Ejection}",
      journal = {\apjl},
         year = 2013,
        month = dec,
       volume = {779},
       number = {2},
          eid = {L29},
        pages = {L29},
          doi = {10.1088/2041-8205/779/2/L29},
       adsurl = {https://ui.adsabs.harvard.edu/abs/2013ApJ...779L..29G}
}

@ARTICLE{gopalswamy2003,
       author = {{Gopalswamy}, N. and {Shimojo}, M. and {Lu}, W. and {Yashiro}, S. and {Shibasaki}, K. and {Howard}, R.~A.},
        title = "{Prominence Eruptions and Coronal Mass Ejection: A Statistical Study Using Microwave Observations}",
      journal = {\apj},
         year = 2003,
        month = mar,
       volume = {586},
       number = {1},
        pages = {562-578},
          doi = {10.1086/367614},
       adsurl = {https://ui.adsabs.harvard.edu/abs/2003ApJ...586..562G}
}

@ARTICLE{green2018,
       author = {{Green}, Lucie M. and {T{\"o}r{\"o}k}, Tibor and {Vr{\v{s}}nak}, Bojan and {Manchester}, Ward and {Veronig}, Astrid},
        title = "{The Origin, Early Evolution and Predictability of Solar Eruptions}",
      journal = {\ssr},
         year = 2018,
        month = feb,
       volume = {214},
       number = {1},
          eid = {46},
        pages = {46},
          doi = {10.1007/s11214-017-0462-5},
archivePrefix = {arXiv},
       eprint = {1801.04608},
 primaryClass = {astro-ph.SR},
       adsurl = {https://ui.adsabs.harvard.edu/abs/2018SSRv..214...46G}
}

@ARTICLE{hebe2004,
       author = {{Cremades}, H. and {Bothmer}, V.},
        title = "{On the three-dimensional configuration of coronal mass ejections}",
      journal = {\aap},
         year = 2004,
        month = jul,
       volume = {422},
        pages = {307-322},
          doi = {10.1051/0004-6361:20035776},
       adsurl = {https://ui.adsabs.harvard.edu/abs/2004A&A...422..307C}
}

@ARTICLE{hebe2006,
       author = {{Cremades}, H. and {Bothmer}, V. and {Tripathi}, D.},
        title = "{Properties of structured coronal mass ejections in solar cycle 23}",
      journal = {Advances in Space Research},
         year = 2006,
        month = jan,
       volume = {38},
       number = {3},
        pages = {461-465},
          doi = {10.1016/j.asr.2005.01.095},
       adsurl = {https://ui.adsabs.harvard.edu/abs/2006AdSpR..38..461C}
}

@ARTICLE{howard2009,
       author = {{Howard}, Timothy A. and {Tappin}, S. James},
        title = "{Interplanetary Coronal Mass Ejections Observed in the Heliosphere: 1. Review of Theory}",
      journal = {\ssr},
         year = 2009,
        month = oct,
       volume = {147},
       number = {1-2},
        pages = {31-54},
          doi = {10.1007/s11214-009-9542-5},
       adsurl = {https://ui.adsabs.harvard.edu/abs/2009SSRv..147...31H}
}

@ARTICLE{illung1985,
       author = {{Illing}, R.~M.~E. and {Hundhausen}, A.~J.},
        title = "{Observation of a coronal transient from 1.2 to 6 solar radii}",
      journal = {\jgr},
         year = 1985,
        month = jan,
       volume = {90},
       number = {A1},
        pages = {275-282},
          doi = {10.1029/JA090iA01p00275},
       adsurl = {https://ui.adsabs.harvard.edu/abs/1985JGR....90..275I}
}

@ARTICLE{jenkins2019,
       author = {{Jenkins}, Jack M. and {Hopwood}, Matthew and {D{\'e}moulin}, Pascal and {Valori}, Gherardo and {Aulanier}, Guillaume and {Long}, David M. and {van Driel-Gesztelyi}, Lidia},
        title = "{Modeling the Effect of Mass-draining on Prominence Eruptions}",
      journal = {\apj},
         year = 2019,
        month = mar,
       volume = {873},
       number = {1},
          eid = {49},
        pages = {49},
          doi = {10.3847/1538-4357/ab037a},
archivePrefix = {arXiv},
       eprint = {1901.10970},
 primaryClass = {astro-ph.SR},
       adsurl = {https://ui.adsabs.harvard.edu/abs/2019ApJ...873...49J}
}

@article{jenkins2021,
  title={Prominence formation by levitation-condensation at extreme resolutions},
  author={Jenkins, Jack and Keppens, Rony},
  journal={\aap},
  volume={646},
  pages={A134},
  year={2021},
  publisher={EDP Sciences}
}

@ARTICLE{jenkins2022,
       author = {{Jenkins}, Jack M. and {Keppens}, Rony},
        title = "{Resolving the solar prominence/filament paradox using the magnetic Rayleigh-Taylor instability}",
      journal = {Nature Astronomy},
         year = 2022,
        month = jul,
       volume = {6},
        pages = {942-950},
          doi = {10.1038/s41550-022-01705-z},
       adsurl = {https://ui.adsabs.harvard.edu/abs/2022NatAs...6..942J}
}

@ARTICLE{johnston2025,
       author = {{Johnston}, Craig D. and {Daldorff}, Lars K.~S. and {Schuck}, Peter W. and {Linton}, Mark G. and {Barnes}, Will T. and {Leake}, James E. and {Daley-Yates}, Simon},
        title = "{Filament Mass Losses Forced by Magnetic Reconnection in the Solar Corona}",
      journal = {\apj},
         year = 2025,
        month = apr,
       volume = {982},
       number = {2},
          eid = {131},
        pages = {131},
          doi = {10.3847/1538-4357/adbae8},
archivePrefix = {arXiv},
       eprint = {2502.18657},
 primaryClass = {astro-ph.SR},
       adsurl = {https://ui.adsabs.harvard.edu/abs/2025ApJ...982..131J}
}

@article{kaneko2017,
  title={Reconnection--condensation model for solar prominence formation},
  author={Kaneko, Takafumi and Yokoyama, Takaaki},
  journal={\apj},
  volume={845},
  number={1},
  pages={12},
  year={2017},
  publisher={IOP Publishing}
}

@article{kaneko2018,
  title={Impact of dynamic state on the mass condensation rate of solar prominences},
  author={{Kaneko}, Ttoto. and {Yokoyama}, Ttoto.},
  journal={\apj},
  volume={869},
  number={2},
  pages={136},
  year={2018},
  publisher={IOP Publishing}
}

@ARTICLE{keppens2025,
       author = {{Keppens}, Rony and {Zhou}, Yuhao and {Xia}, Chun},
        title = "{Modeling multiphase plasma in the corona: prominences and rain}",
      journal = {arXiv e-prints},
         year = 2025,
        month = oct,
          eid = {arXiv:2510.25336},
        pages = {arXiv:2510.25336},
archivePrefix = {arXiv},
       eprint = {2510.25336},
 primaryClass = {astro-ph.SR},
       adsurl = {https://ui.adsabs.harvard.edu/abs/2025arXiv251025336K}
}

@ARTICLE{liu2025,
       author = {{Liu}, Qingjun and {Jiang}, Chaowei and {Liu}, Zhipeng},
        title = "{MHD Simulations of the Slow-rise Phase of Solar Eruptions Initiated from a Sheared Magnetic Arcade}",
      journal = {Research in Astronomy and Astrophysics},
         year = 2025,
        month = may,
       volume = {25},
       number = {5},
          eid = {051002},
        pages = {051002},
          doi = {10.1088/1674-4527/adcb8e},
archivePrefix = {arXiv},
       eprint = {2504.07353},
 primaryClass = {astro-ph.SR},
       adsurl = {https://ui.adsabs.harvard.edu/abs/2025RAA....25e1002L}
}

@ARTICLE{li2025,
       author = {{Li}, Xiaohong and {Zhou}, Yuhao and {Keppens}, Rony},
        title = "{Response of the solar atmosphere to flux emergence: With emergence-driven prominence formation}",
      journal = {\aap},
         year = 2025,
        month = jun,
       volume = {698},
          eid = {A232},
        pages = {A232},
          doi = {10.1051/0004-6361/202553957},
archivePrefix = {arXiv},
       eprint = {2505.03521},
 primaryClass = {astro-ph.SR},
       adsurl = {https://ui.adsabs.harvard.edu/abs/2025A&A...698A.232L}
}

@ARTICLE{lee2017,
       author = {{Lee}, Jin-Yi and {Raymond}, John C. and {Reeves}, Katharine K. and {Moon}, Yong-Jae and {Kim}, Kap-Sung},
        title = "{Heating of an Erupting Prominence Associated with a Solar Coronal Mass Ejection on 2012 January 27}",
      journal = {\apj},
         year = 2017,
        month = jul,
       volume = {844},
       number = {1},
          eid = {3},
        pages = {3},
          doi = {10.3847/1538-4357/aa79a4},
archivePrefix = {arXiv},
       eprint = {1706.09116},
 primaryClass = {astro-ph.SR},
       adsurl = {https://ui.adsabs.harvard.edu/abs/2017ApJ...844....3L}
}

@ARTICLE{DiLorenzo2025,
  author  = {Di Lorenzo, L. and Cremades, H. and López, F. and Balmaceda, L. and Talpeanu, D. C. and D’Huys, E. and Mierla, M. and Lloveras, D. and Aznar Cuadrado, R.},
  title   = {A closer look at Streamer Blowout Coronal Mass Ejections: Multi-wavelength and multi-viewpoint analysis of morphology and kinematics},
  journal = {SoPh},
  year    = {2025},
  note    = {submitted}
}

@ARTICLE{landi2010,
       author = {{Landi}, E. and {Raymond}, J.~C. and {Miralles}, M.~P. and {Hara}, H.},
        title = "{Physical Conditions in a Coronal Mass Ejection from Hinode, Stereo, and SOHO Observations}",
      journal = {\apj},
         year = 2010,
        month = mar,
       volume = {711},
       number = {1},
        pages = {75-98},
          doi = {10.1088/0004-637X/711/1/75},
       adsurl = {https://ui.adsabs.harvard.edu/abs/2010ApJ...711...75L}
}

@ARTICLE{low2001,
       author = {{Low}, B.~C.},
        title = "{Coronal mass ejections, magnetic flux ropes, and solar magnetism}",
      journal = {\jgr},
         year = 2001,
        month = nov,
       volume = {106},
       number = {A11},
        pages = {25141-25164},
          doi = {10.1029/2000JA004015},
       adsurl = {https://ui.adsabs.harvard.edu/abs/2001JGR...10625141L}
}

@article{low2012,
  title={The hydromagnetic interior of a solar quiescent prominence. II. Magnetic discontinuities and cross-field mass transport},
  author={Low, BC and Liu, W and Berger, T and Casini, R},
  journal={\apj},
  volume={757},
  number={1},
  pages={21},
  year={2012},
  publisher={IOP Publishing}
}

@ARTICLE{mccauley2015,
       author = {{McCauley}, P.~I. and {Su}, Y.~N. and {Schanche}, N. and {Evans}, K.~E. and {Su}, C. and {McKillop}, S. and {Reeves}, K.~K.},
        title = "{Prominence and Filament Eruptions Observed by the Solar Dynamics Observatory: Statistical Properties, Kinematics, and Online Catalog}",
      journal = {\solphys},
         year = 2015,
        month = jun,
       volume = {290},
       number = {6},
        pages = {1703-1740},
          doi = {10.1007/s11207-015-0699-7},
archivePrefix = {arXiv},
       eprint = {1505.02090},
 primaryClass = {astro-ph.SR},
       adsurl = {https://ui.adsabs.harvard.edu/abs/2015SoPh..290.1703M}
}

@ARTICLE{priest1986,
       author = {{Priest}, E.~R.},
        title = "{Magnetic Reconnection on the Sun}",
      journal = {Mitteilungen der Astronomischen Gesellschaft Hamburg},
         year = 1986,
        month = jan,
       volume = {65},
        pages = {41},
       adsurl = {https://ui.adsabs.harvard.edu/abs/1986MitAG..65...41P}
}

@ARTICLE{parker1953,
       author = {{Parker}, Eugene N.},
        title = "{Instability of Thermal Fields.}",
      journal = {\apj},
         year = 1953,
        month = may,
       volume = {117},
        pages = {431},
          doi = {10.1086/145707},
       adsurl = {https://ui.adsabs.harvard.edu/abs/1953ApJ...117..431P}
}

@ARTICLE{sen2025,
       author = {{Sen}, Samrat and {Nayak}, Sushree S. and {Antolin}, Patrick},
        title = "{Role of magnetic shear distribution in the formation of eruptive flux ropes}",
      journal = {\aap},
         year = 2025,
        month = nov,
       volume = {703},
          eid = {A241},
        pages = {A241},
          doi = {10.1051/0004-6361/202556232},
archivePrefix = {arXiv},
       eprint = {2509.11416},
 primaryClass = {astro-ph.SR},
       adsurl = {https://ui.adsabs.harvard.edu/abs/2025A&A...703A.241S}
}

@ARTICLE{shibata2001,
       author = {{Shibata}, Kazunari and {Tanuma}, Syuniti},
        title = "{Plasmoid-induced-reconnection and fractal reconnection}",
      journal = {Earth, Planets and Space},
         year = 2001,
        month = jun,
       volume = {53},
       number = {6},
        pages = {473-482},
          doi = {10.1186/BF03353258},
archivePrefix = {arXiv},
       eprint = {astro-ph/0101008},
 primaryClass = {astro-ph},
       adsurl = {https://ui.adsabs.harvard.edu/abs/2001EP&S...53..473S}
}

@article{terradas2015,
  title={Morphology and dynamics of solar prominences from 3D MHD simulations},
  author={Terradas, J and Soler, R and Luna, M and Oliver, R and Ballester, JL},
  journal={\apj},
  volume={799},
  number={1},
  pages={94},
  year={2015},
  publisher={IOP Publishing}
}

@ARTICLE{tripathi2004,
       author = {{Tripathi}, D. and {Bothmer}, V. and {Cremades}, H.},
        title = "{The basic characteristics of EUV post-eruptive arcades and their role as tracers of coronal mass ejection source regions}",
      journal = {\aap},
         year = 2004,
        month = jul,
       volume = {422},
        pages = {337-349},
          doi = {10.1051/0004-6361:20035815},
       adsurl = {https://ui.adsabs.harvard.edu/abs/2004A&A...422..337T}
}

@INPROCEEDINGS{tsurutani2019,
       author = {{Tsurutani}, B. and {Lakhina}, G.~S.},
        title = "{An Extreme Coronal Mass Ejection and Consequences for the Magnetosphere and Earth}",
    booktitle = {AGU Fall Meeting Abstracts},
         year = 2019,
       volume = {2019},
        month = dec,
          eid = {SM13E-3352},
        pages = {SM13E-3352},
       adsurl = {https://ui.adsabs.harvard.edu/abs/2019AGUFMSM13E3352T}
}

@article{xia2016,
  title={Internal dynamics of a twin-layer solar prominence},
  author={Xia, Chun and Keppens, Rony},
  journal={\apjl},
  volume={825},
  number={2},
  pages={L29},
  year={2016},
  publisher={IOP Publishing}
}

@ARTICLE{xia2018,
       author = {{Xia}, C. and {Teunissen}, J. and {El Mellah}, I. and {Chan{\'e}}, E. and {Keppens}, R.},
        title = "{MPI-AMRVAC 2.0 for Solar and Astrophysical Applications}",
      journal = {\apjs},
         year = 2018,
        month = feb,
       volume = {234},
       number = {2},
          eid = {30},
        pages = {30},
          doi = {10.3847/1538-4365/aaa6c8},
archivePrefix = {arXiv},
       eprint = {1710.06140},
 primaryClass = {astro-ph.SR},
       adsurl = {https://ui.adsabs.harvard.edu/abs/2018ApJS..234...30X}
}

@ARTICLE{xing2024,
       author = {{Xing}, Chen and {Aulanier}, Guillaume and {Cheng}, Xin and {Xia}, Chun and {Ding}, Mingde},
        title = "{Unveiling the Initiation Route of Coronal Mass Ejections through Their Slow Rise Phase}",
      journal = {\apj},
         year = 2024,
        month = may,
       volume = {966},
       number = {1},
          eid = {70},
        pages = {70},
          doi = {10.3847/1538-4357/ad2ea9},
archivePrefix = {arXiv},
       eprint = {2402.16679},
 primaryClass = {astro-ph.SR},
       adsurl = {https://ui.adsabs.harvard.edu/abs/2024ApJ...966...70X}
}

@ARTICLE{xing2024b,
       author = {{Xing}, Yaoyu and {Duan}, Aiying and {Jiang}, Chaowei},
        title = "{An explanation for the slow-rise phase of solar eruptions}",
      journal = {\mnras},
         year = 2024,
        month = oct,
       volume = {534},
       number = {1},
        pages = {107-116},
          doi = {10.1093/mnras/stae2088},
       adsurl = {https://ui.adsabs.harvard.edu/abs/2024MNRAS.534..107X}
}

@ARTICLE{xing2025,
       author = {{Xing}, Chen and {Cheng}, Xin and {Aulanier}, Guillaume and {Ding}, Mingde},
        title = "{Initiation Route of Coronal Mass Ejections. II. The Role of Filament Mass}",
      journal = {\apj},
         year = 2025,
        month = jun,
       volume = {986},
       number = {1},
          eid = {37},
        pages = {37},
          doi = {10.3847/1538-4357/adceb5},
archivePrefix = {arXiv},
       eprint = {2504.14876},
 primaryClass = {astro-ph.SR},
       adsurl = {https://ui.adsabs.harvard.edu/abs/2025ApJ...986...37X}
}

@ARTICLE{vale,
       author = {{Sieyra}, M.~V. and {Strugarek}, A. and {Prasad}, A. and {Wagner}, A. and {D{\'e}moulin}, P. and {Moreno-Insertis}, F. and {Finley}, A.~J. and {Joshi}, R. and {Blaise}, A. and {Brun}, A.~S. and {Buchlin}, E.},
        title = "{Formation and rising phase of a flux rope through data-constrained simulations}",
      journal = {\aap},
         year = 2026,
         note = {Submitted}
}

@ARTICLE{vourlidas2013,
       author = {{Vourlidas}, A. and {Lynch}, B.~J. and {Howard}, R.~A. and {Li}, Y.},
        title = "{How Many CMEs Have Flux Ropes? Deciphering the Signatures of Shocks, Flux Ropes, and Prominences in Coronagraph Observations of CMEs}",
      journal = {\solphys},
         year = 2013,
        month = may,
       volume = {284},
       number = {1},
        pages = {179-201},
          doi = {10.1007/s11207-012-0084-8},
archivePrefix = {arXiv},
       eprint = {1207.1599},
 primaryClass = {astro-ph.SR},
       adsurl = {https://ui.adsabs.harvard.edu/abs/2013SoPh..284..179V}
}

@ARTICLE{vourlidas2014,
       author = {{Vourlidas}, Angelos},
        title = "{The flux rope nature of coronal mass ejections}",
      journal = {Plasma Physics and Controlled Fusion},
         year = 2014,
        month = jun,
       volume = {56},
       number = {6},
          eid = {064001},
        pages = {064001},
          doi = {10.1088/0741-3335/56/6/064001},
       adsurl = {https://ui.adsabs.harvard.edu/abs/2014PPCF...56f4001V}
}

@ARTICLE{wang2016,
       author = {{Wang}, Bing and {Chen}, Yao and {Fu}, Jie and {Li}, Bo and {Li}, Xing and {Liu}, Wei},
        title = "{Dynamics of a Prominence-horn Structure during Its Evaporation in the Solar Corona}",
      journal = {\apjl},
         year = 2016,
        month = aug,
       volume = {827},
       number = {2},
          eid = {L33},
        pages = {L33},
          doi = {10.3847/2041-8205/827/2/L33},
archivePrefix = {arXiv},
       eprint = {1608.04095},
 primaryClass = {astro-ph.SR},
       adsurl = {https://ui.adsabs.harvard.edu/abs/2016ApJ...827L..33W}
}

@ARTICLE{webb2012,
       author = {{Webb}, David F. and {Howard}, Timothy A.},
        title = "{Coronal Mass Ejections: Observations}",
      journal = {Living Reviews in Solar Physics},
         year = 2012,
        month = dec,
       volume = {9},
       number = {1},
          eid = {3},
        pages = {3},
          doi = {10.12942/lrsp-2012-3},
       adsurl = {https://ui.adsabs.harvard.edu/abs/2012LRSP....9....3W}
}

@ARTICLE{yuhao2014,
       author = {{Zhou}, Yu-Hao and {Chen}, Peng-Fei and {Zhang}, Qing-Min and {Fang}, Cheng},
        title = "{Dependence of the length of solar filament threads on the magnetic configuration}",
      journal = {Research in Astronomy and Astrophysics},
         year = 2014,
        month = may,
       volume = {14},
       number = {5},
          eid = {581-588},
        pages = {581-588},
          doi = {10.1088/1674-4527/14/5/007},
archivePrefix = {arXiv},
       eprint = {1312.7181},
 primaryClass = {astro-ph.SR},
       adsurl = {https://ui.adsabs.harvard.edu/abs/2014RAA....14..581Z}
}

@ARTICLE{yoshi2025,
       author = {{Yoshihisa}, Takero and {Yokoyama}, Takaaki and {Kaneko}, Takafumi},
        title = "{Conditions for Solar Prominence Formation Triggered by Single Localized Heating}",
      journal = {\apj},
         year = 2025,
        month = jan,
       volume = {978},
       number = {1},
          eid = {94},
        pages = {94},
          doi = {10.3847/1538-4357/ad9908},
archivePrefix = {arXiv},
       eprint = {2411.18193},
 primaryClass = {astro-ph.SR},
       adsurl = {https://ui.adsabs.harvard.edu/abs/2025ApJ...978...94Y}
}

@ARTICLE{zaitsev2018,
       author = {{Zaitsev}, V.~V. and {Stepanov}, A.~V.},
        title = "{Prominence activation by increase in electric current}",
      journal = {Journal of Atmospheric and Solar-Terrestrial Physics},
         year = 2018,
        month = nov,
       volume = {179},
        pages = {149-153},
          doi = {10.1016/j.jastp.2018.06.004},
       adsurl = {https://ui.adsabs.harvard.edu/abs/2018JASTP.179..149Z}
}

@ARTICLE{zhao2017,
       author = {{Zhao}, Xiaozhou and {Xia}, Chun and {Keppens}, Rony and {Gan}, Weiqun},
        title = "{Formation and Initiation of Erupting Flux Rope and Embedded Filament Driven by Photospheric Converging Motion}",
      journal = {\apj},
         year = 2017,
        month = jun,
       volume = {841},
       number = {2},
          eid = {106},
        pages = {106},
          doi = {10.3847/1538-4357/aa7142},
       adsurl = {https://ui.adsabs.harvard.edu/abs/2017ApJ...841..106Z}
}

@ARTICLE{zhao2022,
       author = {{Zhao}, Xiaozhou and {Keppens}, Rony},
        title = "{Plasmoid-fed Prominence Formation (PF$^{2}$) During Flux Rope Eruption}",
      journal = {\apj},
         year = 2022,
        month = mar,
       volume = {928},
       number = {1},
          eid = {45},
        pages = {45},
          doi = {10.3847/1538-4357/ac54a4},
archivePrefix = {arXiv},
       eprint = {2202.08367},
 primaryClass = {astro-ph.SR},
       adsurl = {https://ui.adsabs.harvard.edu/abs/2022ApJ...928...45Z}
}

@ARTICLE{zhou2023,
       author = {{Zhou}, Yuhao and {Li}, Xiaohong and {Hong}, Jie and {Keppens}, Rony},
        title = "{Winking filaments due to cyclic evaporation-condensation}",
      journal = {\aap},
         year = 2023,
        month = jul,
       volume = {675},
          eid = {A31},
        pages = {A31},
          doi = {10.1051/0004-6361/202346004},
archivePrefix = {arXiv},
       eprint = {2305.13237},
 primaryClass = {astro-ph.SR},
       adsurl = {https://ui.adsabs.harvard.edu/abs/2023A&A...675A..31Z}
}
\end{document}